\documentclass[aps,twocolumn,showpacs,preprintnumbers,superscriptaddress]{revtex4-2}
\usepackage{amsmath,amssymb}
\usepackage{comment}
\usepackage{graphicx}
\usepackage{color}
\usepackage[colorlinks,bookmarks=true,citecolor=blue,linkcolor=red,urlcolor=blue]{hyperref}
\usepackage{hyperref}

\def\cL{\mathcal{L}}

\def\cS{\mathcal{S}}

\newcommand\x{\mathbf{x}}
\newcommand\y{\mathbf{y}}
\newcommand\+{\dagger}
\newcommand\<{\langle}
\renewcommand\>{\rangle}
\renewcommand\d{\partial}

\def\d{\partial}
\def\cL{\mathcal{L}}

\def\cS{\mathcal{S}}

\def\>{\rangle}
\def\<{\langle}

\renewcommand\d{\partial}

\newcommand\mk{\mathbf{k}}
\newcommand\mx{\mathbf{x}}
\newcommand\my{\mathbf{y}}
\newcommand\mq{\mathbf{q}}

\begin{document}
	
	
\title{Dirac Composite Fermion Theory of General Jain's Sequences}

	
\author{Dung Xuan Nguyen}
\affiliation{Brown Theoretical Physics Center and Department of Physics,
	Brown University, 182 Hope Street, Providence, RI 02912, USA}
\author{Dam Thanh Son}
\affiliation{Kadanoff Center for Theoretical Physics, University of Chicago, Chicago, Illinois 60637, USA}
	
\begin{abstract}

We reconsider the composite fermion theory of general Jain's
sequences with filling factor $\nu=N/(4N\pm1)$.
We show that Goldman and Fradkin's proposal of a Dirac composite
fermion leads to a violation of the Haldane bound on the coefficient
of the static structure factor.  To resolve this apparent
contradiction, we add to the effective theory a gapped chiral mode
(or modes) which already exists in the Fermi liquid state at
$\nu=1/4$.
We interpret the additional mode as an internal degree of freedom of
the composite fermion, related to area-preserving deformations of the
elementary droplet built up from electrons and correlation holes.  In
addition to providing a suitable static structure factor, our model
also gives the expected Wen-Zee shift and a Hall conductivity that
manifests Galilean invariance.  We show that the charge density in the
model satisfies the long-wavelength version of the
Girvin-MacDonald-Platzman algebra.
\end{abstract}
	
\maketitle    
	
\section{Introduction.}
		
Since the discovery of the fractional quantum Hall effect (FQHE) in
1982 \cite{FQH1,FQH2}, many theoretical models have been invented to
explain different aspects of this fascinating phenomenon.  However,
with no proposed theory capable of explaining all the richness of the
FQHE, the latter remains one of the most non-trivial questions of
condensed matter physics. One breakthrough idea was proposed by Jain
\cite{Jain1}, who suggested that the low energy degree of freedom of
FQHE is the composite fermion (CF), which can be thought of as an
an electron moving together with an even number of magnetic flux quanta,
a picture inspired by previous ideas~\cite{Flux1,Flux2,Flux3}.  The
The composite fermion model has achieved significant success. At the
mean-field level, the composite fermion picture explains the presence
(though not the magnitude) of the FQH gap, mapping the latter to an
integer quantum Hall (IQH) gap of the composite fermions. Inspired by
Jain's CF, Halperin, Lee, and Read introduced \cite{HLR} a field
theory---the HLR theory---that
predicts a Fermi-liquid-like behavior of composite fermions near
half-filling ($\nu=\frac12$), which been confirmed experimentally
\cite{Kang}.

Despite its enormous success, the HLR theory leaves many issues
unresolved.  One issue is the scale of the energy gap, which is
incorrectly predicted to be the cyclotron frequency in the simplest
version of the flux attachment procedure.  This issue can be resolved
by treating the effective mass of the composite fermion as a
phenomenological parameter, and by adding new terms in the effective
Lagrangian, which, in the lowest Landau level limit, amounts to
assigning an electric dipole moment to the composite fermion.  A more
serious problem is the apparent breaking of particle-hole symmetry
(PH), which is the emergent symmetry of the lowest Landau level
(LLL)~\cite{PHKivelson}.  Flux attachment equates the composite
fermion density with the electron density, breaking the symmetry
between electrons and holes, which holds in experiment
\cite{Reflection3}.  The Dirac composite fermion theory
\cite{Son:Dirac} has been proposed as an effective field theory for
FQH states near half-filling, which explicitly incorporates PH
symmetry by assigning a Berry phase of $\pi$ for the CF around the
Fermi surface.  The theory gives reasonable predictions for the
electromagnetic response functions~\cite{Nguyen2018a}.

In contrast to the case of $\nu=1/2$, at $\nu=1/4$ there is no
symmetry that can be used to constrain the Berry phase of the CF.
Recently Goldman and Fradkin, motivated
by the experimentally observed 
reflection symmetry relating the $I-V$ curves of states with filling
fraction $\nu<1/4$ and $\nu>1/4$~\cite{Kang,Reflection2,Reflection3},
proposed a Dirac composite fermion
model for the $\nu=N/(4N+1)$ Jain sequences (and for the general
Jain's sequences with $\nu=\frac{N}{2nN \pm1}$).  The Berry phase of the CF in this model is equal
to $\pi$. The similar effective theory for FQH states near $\nu=1/4$
was proposed by Wang in \cite{Wang:2019} motivated by the numerical study of the many-body Berry phase. 

We will show in Goldman and Fradkin's model the static
structure factor of the $\nu=N/(4N+1)$ [and, more generally,
$\nu=N/(2nN+1)$] states violates the Haldane bound, which places
a lower bound on the coefficient of the leading $q^4$ behavior of the
projected static structure factor \footnote{We will leave an explicit derivation of the Haldane bound to Appendix \ref{sec:HB}} 
\begin{equation}
	s_4 \geq \frac{|\mathcal{S}-1|}{8},
	\end{equation}
where $s_4$ is the coeffient of $q^4 \ell_B^4$ of the projected static structure factor and $\mathcal{S}$ is the Wen-Zee shift. 
This paper aims to construct an effective theory that adequately
describes the physics of the $\nu=N/(4N\pm1)$ Jain's sequences in the
lowest Landau level limit.  We combine the ideas of preceding works
\cite{Son:Dirac, Goldman2018,  Haldane:2009idea, Gromov2017}, including
Haldane's idea of a new geometric degree of freedom of the FQH
system~\cite{Haldane:2009idea} and that of the bimetric theory of this
degree of freedom~\cite{Gromov2017}.  Our model is a hybrid model
that includes a Dirac composite fermion sector and an additional
spin-2 mode with the effective action of the form of the bimetric
theory.
In this model, in addition to the low-energy GMP mode at the
energy scale of the effective cyclotron energy, which tends to zero as
$N\to\infty$, there is an extra high-energy mode with energy which
remains finite in the limit $N\to\infty$, and hence must already exist
in the $\nu=1/2n$ Fermi-liquid state.  With the extra mode, which we
call the Haldane mode, the issue with the Haldane bound of the projected
static structure factor in the previous Dirac composite fermion model
\cite{Goldman2018} can be fixed.  We derive the topological Wen-Zee
shift of Jain's sequences that matches the expected results in the
literature \cite{Cho:WZJain1,Gromov:WZJain2}.  We also evaluate the
wave-number dependence of the Hall conductivity and verify its
connection with the Hall viscosity~\cite{Son-Hoyos}. We then rederive
the long-wavelength limit of GMP algebra as evidence that the
theory is appropriate for the LLL limit.

The plan of the paper is as follows. In Sec.\ \ref{sec:EFT} we propose
our effective theory, introducing its ingredients and briefly
discussing the motivations. Section \ref{sec:WZ} is devoted to
calculation of the Wen-Zee shift, the Hall viscosity,
and the Hall conductivity. In
Sec.\ \ref{sec:GMP}, we derive the long-wavelength version of GMP
algebra. In Sec.\ \ref{sec:SSF} we calculate the projected static
structure factor using a semiclassical approach developed in
Refs.~\cite{Nguyen2018a,Nguyen2018b} and comment on the importance of the
extra mode. Lastly, we conclude our paper and discuss the open
questions in Sec.\ \ref{sec:concl}.  Some technical details of the
calculations are left to the Appendices.

\section{Effective field theory}
\label{sec:EFT}

Inspired by previous works
\cite{Goldman2018,Haldane:2009idea,Haldane2011,Gromov2017}, we propose
an effective theory for the Jain sequences around $\nu=1/(2n)$,
$\nu_{\pm}=\frac{N}{2n N \pm 1}$, where $n$ and $N$ are integers.  In
practice, only Jain sequences have been seen only around $\nu=1/2$ and
$\nu=1/4$, so in practice for $n=2$, though for generality we will keep
$n$ arbitrary in our formulas.

As we will see in Section \ref{sec:SSF}, the projected static
structure factor for the $\nu=N/(4N+1)$ calculated in the Dirac
composite fermion model by Goldman and Fradkin \cite{Goldman2018}
violates the Haldane bound.  In order to fix this issue, we add an
extra chiral mode to the model.  Though the microscopic nature of this
mode is not important for our effective theory, one can, following
Haldane~\cite{Haldane:2009idea,Haldane2011}, think about this mode as
corresponding to the area-preserving deformation of the composite
object (``elementary droplet'') built up from electrons and
correlation holes.  In this paper, we show that with the proper choice
of the coupling of the extra mode with external background
electromagnetic field, we can reproduce the known physical
results for the general Jain's sequence.

\subsection{Symmetry of the Lowest Landau Level}

The system of electrons with Land\'e factor
$\mathfrak{g}=2$~\footnote{The Land\'e factor is introduced for
	convenience and does not change physical quantities computed in this
	paper, for example, the static structure factor.} in a constant
magnetic field $B$ has a well defined lowest Landau level (LLL) limit
where the electron mass goes to
zero~\cite{Son2013,Geracie:2015}. Inheriting the symmetry of
non-relativistic massless fermion with $\mathfrak{g}=2$
\cite{Son2013,prabhu2017electrons} \footnote{We take $\mathfrak{g}=2$ only for convenience in which the mathematical treatment of the lowest Landau level limit is simplified. A different value of $\mathfrak{g}$ only modifies chemical potential by a constant value \cite{Geracie:2015} that doesn't change the physical conclusions in this paper.}, the external gauge field appears
in the effective action as the combination
\begin{equation}
	\label{eq:Amu}
	\tilde{A}_0=A_0-\frac{1}{2}\epsilon^{ij}\d_i v_j+\frac{1}{2}\omega_0, \qquad \tilde{A}_i=A_i+\frac{1}{2}\omega_i.
\end{equation}
where $v^i$ is the drift velocity, given by
$v^i=\frac{\epsilon^{ij}E_j}{B}$ at the leading order, and
$\omega_\mu$ is the spin connection of the Newton-Cartan space that
was defined through derivative of vielbein $e^a_i$ \cite{Son2013}
\begin{align}
	\omega_0 &=\frac{1}{2}\epsilon^{ij}\d_i v_j +\frac{1}{2}\epsilon^{ab} e^{aj}\d_0 e^{b}_j,\\ \omega_i &= \frac{1}{2}\epsilon^{ab} e^{aj}\d_i e^{b}_j-\frac{1}{2}\epsilon^{jk}\d_j g_{ik}.
\end{align}
In particular, in flat space and uniform magnetic field
\begin{equation}
	\label{eq:A0}
	\tilde{A}_0=\left(1+\frac{\nabla^2}{4B}+\cdots\right)A_0.
\end{equation}
Effective theories on the LLL, including composite fermion models, should
couple with the external gauge field through $\tilde A_\mu$.

\subsection{Composite fermion sector}

In this paper, we propose the following effective action of Jain's sequences
\begin{align}
		\label{eq:action}
		S=S_{CF}+S_{\text{Haldane}},
\end{align}
where the Dirac composite fermion action is given by 
\begin{multline}\label{eq:LCF}
  S_{CF} = \int\! d^3 x\, \sqrt{g}\frac{i}{2}\Bigl(
  \psi^\dagger\overleftrightarrow{D}_{\!\!t} \psi
  + e^i_a \psi^\dagger \sigma^a \overleftrightarrow{D}_{\!\!i} \psi+v^i \psi^\dagger \overleftrightarrow{D}_{\!\!i} \psi\Bigr)\\
  \int \left[ -\frac{1}{8\pi}\left(1-\frac{1}{n}\right)a da - \frac{1}{4\pi n} \tilde{A}da + \frac{1}{8\pi n}\tilde{A}d \tilde{A} \right],
\end{multline}
where $\sigma^a$ is the Pauli matrix and $a_\mu$ is the emergent gauge
field.  As in Ref.~\cite{Goldman2018}, the Dirac composite fermion is
massless.  Here we chose the Fermi velocity of composite fermion
$v_F=1$ to simplify the formulae \cite{Nguyen2018a}; the main results
of this paper do not depend on $v_F$.  The covariant derivative of the
composite fermion is
\begin{equation}
   D_\mu\psi \equiv (\d_\mu-i a_\mu+\frac{i}{2}\sigma^3 \omega_\mu)\psi .
\end{equation}
The CF is charged neutral without direct coupling to the external
electromagnetic field.  The third term of \eqref{eq:LCF} implies that
composite fermion has a dipole moment perpendicular to its momentum.
This term is new compared to previous work on Dirac composite fermion
model with the filling fraction is near $1/2n$
\cite{Goldman2018,Wang:2019}. The neutrality of the composite fermions
, as well as its dipole moment, are features of the Dirac composite
fermion near half-filling \cite{Son:Dirac, Nguyen2018a}.

Differentiating the action over $a_0$ gives us a constraint (we work
in flat space)
\begin{equation}\label{eq:rhoCF}
	\bar{\rho}_{\text{\tiny{CF}}}=\frac{1}{4\pi}\left(1-\frac{1}{n}\right)\bar{b}+\frac{1}{4\pi n} B,
\end{equation}
while differentiating the action with respect to $A_0$ and taking
average over space gives the average physical charge density,
\begin{equation}
	\label{eq:rhoe}
	\bar{\rho}_e=\frac{1}{4\pi n} B-\frac{1}{4\pi n}\bar{b}.
\end{equation}
The Haldane mode will not change the above equation; physically, the area-preserving deformations of the elementary droplets do not change the
average density.  
The specific values of emergent magnetic field
\begin{equation}\label{eq:bbar}
	\bar{b}=\pm\frac{1}{2nN\pm 1}B,
\end{equation}
correspond to FQH filling fractions of the Jain states 
\begin{equation}
	\nu_{\pm}=\frac{N}{2n N \pm 1},
\end{equation}
where the filling fraction of composite fermion is 
\begin{equation}
	\label{eq:nuCF}
	\nu^{CF}_{\pm}=\pm N + \frac{1}{2}.
\end{equation}
which are the filling factors of IQH states of a massless Dirac
fermion.

\subsection{Haldane sector}
	
The Haldane mode (named so because it embodies Haldane's idea of an
emergent degree of freedom of the
FQHE~\cite{Haldane:2009idea,Haldane2011}) is a gapped dynamical spin-2
degree of freedom described by the extra metric $\hat{g}_{ij}$.  We
define the emergent vielbein as the square root of the ambient metric
\begin{equation}
	\hat{g}_{ij}=\hat{e}^\alpha_i \hat{e}^\beta_j \delta_{\alpha\beta},
\end{equation}  
and we also denote the matrix inverse of $\hat{e}^\alpha_i$ by
$\hat{E}^i_\alpha$. The compatibility conditions,
\begin{align}
		\hat{\nabla}_k \hat{g}_{ij}&=\d_k \hat{g}_{ij}-\hat{\Gamma}^{i}_{k,i}\hat{g}_{ij}-\hat{\Gamma}^{l}_{k,j}\hat{g}_{lj}=0,\\
		\hat{\nabla}_\mu \hat{e}_\nu^A&=\d_\mu \hat{e}_\nu^A-\hat{\Gamma}^\lambda_{\nu,\mu}\hat{e}^A_\lambda+\hat{\omega}^A_{B,\mu}\hat{e}^B_\nu=0,
\end{align}
are satisfied by 
\begin{align}
		\hat{\omega}_0&=\frac{1}{2}\epsilon^{\alpha\beta} \hat{E}^i_\beta \d_0 \hat{e}^\alpha_i,\\
		\hat{\omega}_j&=\frac{1}{2}\epsilon^{\alpha\beta}\left( \hat{E}^i_\beta \d_j \hat{e}^\alpha_i- \hat{\Gamma}^k_{i,j}\hat{e}^\alpha_k \hat{E}^i_\beta \right),\\
		\hat{\Gamma}^{i}_{k,j}&=\frac{1}{2}\hat{G}^{il}\left(\d_j \hat{g}_{kl}+\d_k \hat{g}_{jl}-\d_l \hat{g}_{jk} \right),\\
		\hat{\Gamma}^i_{j,0}&=\frac{1}{2}\hat{G}^{ik}\d_0 \hat{g}_{jk}.
\end{align}
We further introduce the definition of the emergent Ricci curvature of  $\hat{g}_{ij}$ as
\begin{equation}
		\hat{R}=\frac{2}{\sqrt{\hat{g}}}(\d_1 \hat{\omega}_2-\d_2 \hat{\omega}_1).
\end{equation}
Restricting the dynamical part of the
the emergent metric to fluctuations that preserve the area, setting
$\hat{g}=g$, we can rewrite the emergent Ricci curvature as 
\begin{equation}
	\hat{R}=\frac{2}{\sqrt{g}}(\d_1 \hat{\omega}_2-\d_2 \hat{\omega}_1).
\end{equation}
We propose to use as the action of the spin-2 mode the leading (in
derivatives) terms of bimetric theory introduced in
Refs.~\cite{Gromov2017,Nguyen2018}.
\begin{multline}\label{eq:Haldane}
	S_{\text{Haldane}}=\int\! \frac{\zeta}{4n\pi}A d\hat{\omega}\\
		-\!\int\! d^3 x\, \sqrt{g}\biggl[\frac{\tilde{m}}{2}\Bigr(\frac{1}{2}\hat{g}_{ij}g^{ij}-\gamma \Bigl)^2+\frac{\zeta}{8n\pi B}g^{ij}\left(\d_i E_j\right)\mathcal{B}\biggr],
\end{multline}
with $\nu$ is the FQH filling fraction, and the magnetic field in
curved space is defined as $\mathcal{B}=\frac{\partial_1
	A_2-\partial_2 A_1}{\sqrt{g}}$. The mass $\tilde{m}>0$ in the
potential term sets the energy gap for the extra spin-2 mode. We
require $\gamma<1$ so that the ground state is
isotropic~\cite{Gromov2017,Nguyen2018}. The last term comes from the
symmetry of LLL, where the temporal component of the emergent spin
connection enters the Lagrangian through the combination
$\hat{\omega}_0+\frac{1}{2}\epsilon^{ij}\d_i v_j$, hence its
coefficient is fixed by that of the first term.  This last term
contributes to the finite-wave-number Hall conductivity at the $k^2$
order.  Besides $\tilde m$ there is only one additional parameter,
$\zeta$ \cite{Gromov2017}. In the subsequent section, we will show
that one can fix the coefficient $\zeta$ for general Jain's sequences
using the Wen-Zee shift.  We will find that $\zeta$ only depends on
$n$ but not $N$, as expected.

\section{Wen-Zee shift $\mathcal{S}$ and Galilean invariant}\label{sec:WZ}

In this section, we find the value of $\zeta$. We show that, with $\zeta=n-1$, one can obtain the expected Wen-Zee shift, the
finite-wave-number Hall conductivity and the Hall viscosity for the
general Jain's sequences $\nu_\pm$. The results demonstrate that the
topological Wen-Zee shift as well as the relationship between the
finite-wave-number Hall conductivity and the shift~\cite{Son-Hoyos}
are reproduced successfully in our model.
	
\subsection{Wen-Zee shift $\mathcal{S}$}
\subsubsection{Filling fraction $\nu_+=\frac{N}{2n N+1}$}
	
At the filling fraction $\nu_+$, the filling fraction of composite
fermion is $\nu^{CF}=N+1/2$. The Dirac composite fermion then forms an
IQH state. One then integrate out the fermion fields $\psi$ and obtain
the Chern-Simons (CS) action
\begin{equation}
	\label{eq:CS}
	\frac{N+1/2}{4\pi}a d a+ \frac{N(N+1)}{4\pi} a d\omega.
\end{equation}
The first term encodes the Hall conductivity of IQH state of the Dirac
fermion~\footnote{Which was also found in the IQH experiment of graphene
	after dividing the experimental result of $\sigma_{xy}$ by the total number of spins and valleys \cite{Gusynin:2005}.}. The second term
represents the effective coupling of Dirac composite fermions with
the background curvature, which reproduces the known Wen-Zee shift of
Dirac IQH states~\cite{Golkar:2014,prabhu2017electrons} \footnote{The similar effective action of non-relativistic IQH states can be found in \cite{GromovIQH}}. We have the
following leading terms in the effective Lagrangian of composite
fermion sector \footnote{Note that, due to the additional dipole term,
	one need to shift $a_0\rightarrow a_0+\frac{1}{2}\frac{k_F}{v_F}v_i
	v^i$ and $a_i \rightarrow a_i -\frac{k_F}{v_F}v_i$ as explained in
	\cite{Nguyen2018a}.  However, the extra terms are higher derivative and
	do not modify the results of this paper.}
\begin{multline}
	\cL_{CF}=\frac{N+1/2}{4\pi} a d a + \frac{N(N+1)}{4 \pi}a d\omega -\frac{1}{8\pi}\left(1-\frac{1}{n}\right)a da \\ -\frac{1}{4\pi n} \tilde{A}da+\frac{1}{8\pi n}\tilde{A}d \tilde{A}.
\end{multline}
We then integrate out the emergent gauge field $a_\mu$ to obtain the final effective action of external background fields 
\begin{multline}
	\label{eq:EFCF1}
	\cL_{CF}=-\frac{n N^2(N+1)^2}{8\pi(2n N+1)} \omega d \omega+\frac{\nu_+}{4\pi}(N+1) \tilde{A} d \omega \\+ \frac{\nu_+}{4\pi}\tilde{A} d\tilde{A}.
\end{multline}
From Eq.~\eqref{eq:EFCF1} we obtain the contribution to the charge
density from the Dirac composite sector in a static background without
electric field \footnote{Due to the definition of $\omega_0$, there is
	an extra contribution to the charge density with the form
	$\frac{1}{\sqrt{g}}\partial_i \partial^i (\cdots)$. However, this term
	will not contribute to the Wen-Zee shift on a closed manifold.}
\begin{equation}\label{eq:barrho1}
	\bar{\rho}^{(1)}=\frac{\nu_+}{2\pi}\mathcal{B}+\frac{\nu_+(N+2)}{8\pi}R,
\end{equation} 
with $R\equiv\frac{2}{\sqrt{g}}(\d_1 \omega_2 -\d_2 \omega_1)$ being
the Ricci curvature of the background metric $g_{ij}$. The first term
of \eqref{eq:barrho1} comes from the electromagnetic CS term
$\tilde{A} d \tilde{A}$. The second term receives contributions from
the mixed CS term $\tilde{A} d \omega$ and the $\tilde{A} d \tilde{A}$
tern due to the spin connection in the definition of
$\tilde{A}$ \footnote{Note the definition of $\tilde{A}$ in
	Eq.~\eqref{eq:Amu}}.

Now let us consider the response of the Haldane sector to background
fields.  At zero frequency, in the isotropic phase, the potential term
in \eqref{eq:Haldane} constraints the ambient metric to follow the
background metric $\hat{g}_{ij}=g_{ij}$~\cite{Gromov2017}.  As a consequence, one can read off the contribution to charge density of
the Haldane sector in a static background without an electric
field,
\begin{equation}\label{eq:barrho2}
  \bar{\rho}^{(2)}=\frac{ \zeta}{8n\pi}\hat{R}=\frac{ \zeta}{8n\pi}R.
\end{equation}
Combining Eqs. \eqref{eq:barrho1} and \eqref{eq:barrho2}, we obtain
the charge density,
\begin{equation}\label{eq:WZ1}
  \rho_e=\frac{\nu_+}{2\pi}\mathcal{B}+\frac{\nu_+(N+2)}{8\pi}R +\frac{\zeta}{8n\pi}R,
\end{equation}
and from it the total charge on a closed manifold 
\begin{equation}
	N_e=\nu_+ \left[N_{\phi}+(N+2+2\zeta)\frac{\chi}{2}\right]+O\left( \frac1N\right),
\end{equation}
where $N_\phi$ is the total number of flux quanta $N_{\phi}=\frac{1}{2\pi}\int
\sqrt{g} \mathcal{B}$, and $\chi$ is the Euler character of the Riemann surface
\begin{equation}
	\chi=\frac{1}{4\pi} \int \sqrt{g} R,
\end{equation}
which means that the Wen-Zee shift \cite{Wen-Zee} is
\begin{equation}
  \cS_{\nu_+}=N+2+2\zeta, 
\end{equation}
where we dropped the term of subsub leading in $1/N$ expansion.

\subsubsection{Filling fraction $\nu_-=\frac{N}{2n N-1}$}
	One can repeat the calculations in for $\nu_+$ and obtain the effective Lagrangian of composite fermion sector for $\nu_-$
	\begin{multline}
	\label{eq:EFCF2}
	\cL_{CF}=-\frac{n N^2(N-1)^2}{8\pi(2n N-1)} \omega d \omega+\frac{\nu_-}{4\pi} (-N+1)\tilde{A} d \omega \\+ \frac{\nu_-}{4\pi}\tilde{A} d\tilde{A}.
	\end{multline}
	We again combine the charge density of CF sector and Haldane sector and obtain
	\begin{equation}
	\label{eq:WZ2}
	\rho_e=\frac{\nu_-}{2\pi}\mathcal{B}+\frac{\nu_-(-N+2)}{8\pi}R+\frac{\zeta}{8n\pi}R,
	\end{equation}
	The first two terms of \eqref{eq:WZ2} come from the effective Lagrangian of composite fermion sector \eqref{eq:EFCF2},  the last term of  \eqref{eq:WZ2} come from the Haldane sector \eqref{eq:Haldane}. We then obtain the Wen-Zee shift, up to leading and first subleading orders in the $1/N$ expansion.
\begin{equation}
	\cS_{\nu_-}=-N+2+2\zeta. 
\end{equation}
We now see that if we chose
\begin{equation}
	\zeta=n-1,
\end{equation}
we can reproduce the expected Wen-Zee shift for both $\nu_+$ and
$\nu_-$ sequences \cite{Cho:WZJain1,Gromov:WZJain2}
\begin{equation}
	\cS_{\nu_+}=N+2n, \qquad \cS_{\nu_-}=-N+2n.
\end{equation}
        
For Jain's sequences near half-filling with $n=1$, the coefficient
$\zeta=0$.  In this case, there is no need for a Haldane mode.  We
argue that, due to the particle-hole symmetry at $\nu=1/2$, the
electrons and holes are uniformly distributed in the elementary
droplets, so their area-preserving deformations do not change the
local charge density and then decoupled from the external
electromagnetic field.
	\subsection{Hall viscosity}
To derive the zero frequency Hall viscosity, one needs to
work out the coefficient of the mixed CS term $A d\omega$ in the
effective action. This coefficient is determined by the Wen-Zee shift
as $\nu \mathcal{S} A d\omega$ \cite{Wen-Zee}. One can also do the
calculation explicitly using the effective Lagrangian of the CF sector
and replace, in the Haldane sector, the emergent spin connection
$\hat{\omega}$ by the spin connection of the background metric
$\omega$ at low energy. We found that
\begin{equation}
	\eta_{0}=\bar{\rho}_e \frac{\cS}{4}
\end{equation}
for both sequences $\nu_-$ and $\nu_+$. 


\subsection{Finite-wave-number Hall conductivity}
In a flat background where $g_{ij}=\delta_{ij}$, using the definitions
of $\omega_0$ and $\tilde{A}$, we can rewrite the CF action
\eqref{eq:EFCF1} as
\begin{equation}
		\cL_{CF}=-\frac{\nu_+}{8\pi B}N(\vec{\nabla}\times \vec{A}) \d_i E_i +\frac{\nu_+}{4\pi}A d A.
\end{equation}
The Lagrangian of the Haldane sector for Jain's sequence $\nu_+$ state
at low energy
can be rewritten as 
\begin{equation}
	\cL_{\text{Haldane}}=-\frac{ \zeta}{8n\pi B }(\vec{\nabla}\times \vec{A}) \d_i E_i. 
\end{equation}
One can derive the current in flat space, with $\zeta=n-1$, as a
function of the applied electric field (we work up to leading and
first subleading order in the $1/N$ expansion).
\begin{equation}
	J^i=\frac{\nu_+}{2\pi}\epsilon^{ij}E_j-\frac{\nu_+}{4\pi B} (\mathcal{S}_{\nu_+}-\mathfrak{g})\epsilon^{ij}\d_j(\vec{\nabla}\cdot \vec{E}),
\end{equation}
where we have assumed the Land\'e factor of electron is
$\mathfrak{g}=2$.  We then obtain the Hall conductivity at finite
momentum
\begin{equation}\label{eq:sigma+}
	\sigma_H^{\nu_+}(q)=\frac{\nu_+}{2\pi}\left(1+\frac{\mathcal{S}_{\nu_+}-\mathfrak{g}}{4}q^2 \ell_B^2\right),
\end{equation}
with the magnetic length is $\ell_B=1/\sqrt{B}$. One can perform the same calculation to yield the finite momentum Hall conductivity of Jain sequence $\nu_-$	
\begin{equation}\label{eq:sigma-}
	\sigma_H^{\nu_-}(q)=\frac{\nu_-}{2\pi}\left(1+\frac{\mathcal{S}_{\nu_-}-\mathfrak{g}}{4}q^2 \ell_B^2\right).
\end{equation}
The wave-number dependence of the Hall conductivity of both sequences
$\nu_+$ and $\nu_-$ satisfy the relationship with the Wen-Zee shift
\cite{Son-Hoyos}. The results imply that our model satisfies the
symmetries of the LLL, as designed.

\section{The GMP algebra}\label{sec:GMP}
In the classic paper \cite{GMP:1986}, Girvin, MacDonald, and Platzman
showed that the LLL projected density operator $\rho_e(\mk)$ obeys the
following (GMP) algebra
\begin{equation}
	[\rho(\mk),\rho(\mq)]=2i e^{\frac{1}{2}(\mk\cdot \mq)\ell_B^2}\sin \left(\frac{\mk \times \mq}{2}\ell_B^2\right)\rho(\mk+\mq).
\end{equation}
The GMP algebra was discovered independently in string theory with the
name $W_{\infty}$ \cite{AVAN:1991,GERASIMOV:1991}. In the FQHE, the
GMP algebra is a consequence of charge conservation and the LLL
constraint.  In the long wave length approximation, the GMP algebra is
reduced to the algebra of area-preserving diffeomorphisms
\cite{ISO:1992}
\begin{equation}
  [\rho(\mk),\rho(\mq)]\approx i(\mk\times\mq)\ell_B^2 \rho(\mk+\mq).
\end{equation}
Near half-filling, the long wavelength limit of GMP algebra has been
shown to arise from Dirac composite fermion theory in
Ref.~\cite{Nguyen2018a}.  In this section, we generalize the
derivation of the GMP algebra for the general Jain sequences.  From
the action \eqref{eq:action}, one can read off the charge density operator
in the flat background where $g_{ij}=\delta_{ij}$, $\hat{R}=2
\epsilon^{ij}\d_i \hat{\omega}_j$ up to subleading order in the
momentum expansion,
\begin{equation}
  \rho_e=\rho^{(1)}+\rho^{(2)},
\end{equation}
where $\rho^{(1)}$ and $\rho^{(2)}$ come from the CF and the Haldane
sectors, respectively.  Explicitly,
\begin{align}
  \rho^{(1)} &= \frac{B-b}{4\pi n}- \frac{\epsilon^{ij}}{B} \d_i \Pi_{j},
     \label{eq:chargeden}\\
  \rho^{(2)} &=\frac{\zeta}{8n\pi}\hat{R}.
\end{align}
The second term of Eq.~\eqref{eq:chargeden} implies that the composite
fermion has a dipole moment perpendicular and proportional to
its momentum.
Employing the canonical equal-time anti-commutation relation of the CF field, 
\begin{equation}
	\label{eq:commutepsi}
	\{\psi(\mx),\, \psi^\dagger(\my) \}=\delta(\mx-\my),
\end{equation}
and the commutation relation of emergent gauge field due to the Chern-Simon term of $a_\mu$ in the action \eqref{eq:LCF}
\begin{equation}
	\label{eq:commuteai}
	[a_i(\mx),a_j(\my)]=-i\frac{4\pi n}{n-1}\epsilon^{ij}\delta(\x-\y),
\end{equation}
one can derive the algebra up to lowest order in  momentum of $\rho^{(1)}$
\begin{equation}
	\label{eq:GMPCF}
	[\rho^{(1)}(\mk),\, \rho^{(1)}(\mq)] = i (\mk \times \mq)\ell^2_B\rho^{(1)}(\mk+\mq).
\end{equation}
Interested reader can find the detailed derivation of
Eq.~\eqref{eq:GMPCF} in Appendix \ref{sec:GMPCF}.  Equation
\eqref{eq:GMPCF} implies the GMP algebra of the CF sector.

Following Ref.~\cite{Gromov2017}, one find the charge density in the Haldane sector also satisfies the GMP algebra,
\begin{equation}
\label{eq:GMPH}
	[\rho^{(2)}(\mk),\rho^{(2)}(\mq)]=i (\mk \times \mq)\ell^2_B\rho^{(2)}(\mk+\mq).
\end{equation}
Combining equations \eqref{eq:GMPCF} and \eqref{eq:GMPH}, we arrive at
the GMP algebra of the electron density operator at the lowest order
in momentum.
\begin{equation}
	\label{eq:GMP}
	[\rho_e(\mk),\rho_e(\mq)]= i (\mk \times \mq)\ell^2_B\rho_e(\mk+\mq).
\end{equation}
The GMP algebra is the key feature of a LLL effective theory.  For the
algebra, it is crucial for the Dirac composite fermion to have an
electric dipole moment~\cite{Nguyen2018a}.  The dipole moment term in
Eq.~\eqref{eq:LCF} thus not only preserves the LLL invariance of the
DCF action but also plays an essential role in the GMP algebra
\eqref{eq:GMP}.
	

	
\section{Dynamics of the CF surface and the static structure factor}
\label{sec:SSF}

In the theory of Dirac composite fermion near half-filling, the
dynamics of the composite fermion sector at low energy can be
interpreted as the deformation of the composite Fermi surface
\cite{Golkar2016,Nguyen2018a,Nguyen2018,Nguyen2018b}. One can apply
the semiclassical equation of motion to analyze the physics at the
long-wavelength limit \footnote{It has been shown that the
	the semiclassical approach gives the same results of electromagnetic
	response as the random phase approximation
	calculation~\cite{NguyenLLEM}.}. In this section, we will extend the
semiclassical approach for Dirac composite fermion theory of general
the Jain's sequences and derive the static structure factor (SSF).

\subsection{Semiclassical equation of motions}

Near the filling fraction $\nu=\frac{1}{2n}$, the Dirac composite
fermion forms the Fermi surface with a small effective background
magnetic field $\bar{b}$ induced by the emergent gauge
$a_\mu$. Following
Refs.~\cite{Golkar2016,Nguyen2018a,Nguyen2018b,Nguyen2018}, we
introduce the deformation of composite Fermi surface at a given
position $\mx$
\begin{equation}
	k_F(\theta,\mx)=k_F+\sum_{m=-\infty}^{\infty}u_m(\mx) e^{-i m \theta},
\end{equation}
where $u_n(\mx)$ is a space dependent functions and $\theta$ is the polar angle in the momentum space. 
The Fermi momentum is related with the composite fermion density by Luttinger's theorem \cite{Luttinger:1960}
\begin{equation}
	\label{eq:kF}
	\frac{k_F^2}{4\pi}=\bar{\rho}_{\text{\tiny{CF}}}=\frac{1}{4\pi}\left(1-\frac{1}{n}\right)\bar{b}+\frac{1}{4\pi n} B.
\end{equation}
$u_1(\x)$ and $u_{-1}(\x)$ can be regard as the complex components of
a vector $u_i(\x)$,
	\begin{equation}
	u_1=u_{\bar{z}}, \qquad u_{-1}=u_z,
	\end{equation}
	where we defined the complex coordinate 
	\begin{equation}
		z=x+iy, \qquad \bar{z}=x-iy,
	\end{equation}
and the complex derivative
\begin{equation}
	\d_z=\frac{1}{2}(\d_x - i \d_y ), \qquad \d_{\bar{z}}=\frac{1}{2}(\d_x + i \d_y ).
\end{equation}
The composite fermion density and momentum operator in this
essentially semiclassical formalism is \cite{Nguyen2018a,Nguyen2018b}
\begin{equation}\label{eq:semiope}
	\rho_{\text{\tiny{CF}}}= \bar{\rho}_{\text{\tiny{CF}}}+\frac{k_F}{2\pi}u_0, \qquad \Pi_i =\frac{k_F^2}{4\pi} u_i.
\end{equation}
The $u_n$ modes satisfies the following commutation relation \cite{Golkar2016,Nguyen2018b} 
\begin{multline}\label{eq:uncomm}
	[u_m(\x),\,  u_{m'}(\x')] =\\ \frac{2\pi}{k_F}\Bigl(
	\frac{m \bar{b}}{k_F} \delta_{m+m',0}  - i\delta_{m+m',1} \d_{\bar z}
		-i \delta_{m+m',-1} \d_z \Bigr) \delta(\x-\x') .
\end{multline}
The same commutation relation for Fermi surface in a magnetic field was
derived by Haldane using bosonization~\cite{Haldane:1994}.  The
Fermi-liquid kinetic equation can be interpreted as the equations of
motion that followed from the commutation relation \eqref{eq:uncomm}
and the Hamiltonian \cite{NguyenLLEM,Nguyen2018a}
\begin{equation}
	\label{eq:Ham}
	H=\frac{v_F k_F}{4\pi}\sum_{m=-\infty}^\infty \int d \x (1+F_m) u_m(\x)u_{-m}(\x) -a_0 \rho_{\text{\tiny{CF}}},
\end{equation}
where $F_m$ are the Landau Fermi-liquid parameters. The Heisenberg
equation of motion of $u_m(\x)$ is \footnote{One can derive the same
equation of motion using the Boltzmann kinetic equation
\cite{NguyenLLEM,Nguyen2018a}.}
\begin{multline}
	\label{eq:eom}
	\dot{u}_m=-i(1+F_m)\frac{\bar{b}v_F}{k_F}u_m-v_F(1+F_{m-1})\d_{\bar z}u_{m-1}\\
	-v_F(1+F_{m+1})u_{m+1}+\delta_{m,1}e_{\bar z}+\delta_{m,-1} e_z,
\end{multline}
with $e_i=\d_i a_0 - \d_0 a_i$. For $m=0$, Eq.~\eqref{eq:eom} gives
\begin{equation}
	\label{eq:u0}
	\dot{u}_0=-v_F(1+F_1) (\d_{\bar z}u_{-1}+ \d_z u_{1}),
\end{equation}
which can be understood as the conservation law of CF number:
$\d_t\rho_{\text{\tiny{CF}}}+\mathbf{\nabla}\cdot
\mathbf{j}_{\text{\tiny{CF}}}=0 $, with CF number current
\begin{equation}
  j_i^{\text{\tiny{CF}}}=\frac{v_F(1+F_1)}{k_F}\Pi_i \,. 
\end{equation}
We will use this semiclassical approach to derive the density-density
correlation function, and specifically the static structure factor
to leading order in the momentum expansion.
\subsection{Projected Static Structure Factor}

In this section, we will use the semiclassical approach to calculate
the leading order in gradient expansion of the projected static
structure factor.  We will show that the extra Haldane mode is crucial
to satisfy the Haldane bound, and the fact that the latter is
satisfied provides a nontrivial check for our proposal.

We combine Eqs.~\eqref{eq:chargeden} and \eqref{eq:semiope} to obtain
the charge density operator of the CF sector in the semiclassical
formalism,
\begin{equation}\label{eq:semi-rho}
	\rho^{(1)}=\frac{B-b}{4\pi n}+\frac{i k_F^2}{2\pi B}\left(\d_z u_1-\d_{\bar z} u_{-1} \right).
\end{equation}
We take the time derivative of above equation and use the equation of
motion \eqref{eq:eom} for $m=0, \pm 1, \pm 2$ to obtain, up to leading
order in the gradient expansion.
\begin{equation}
	\label{eq:rhodot}
	\dot{\rho}^{(1)}=\frac{\dot{b}}{4 \pi}\left(\frac{k_F^2}{B}-\frac{1}{n}-\frac{\bar b}{B}\frac{n-1}{n}\right)+\frac{k_F^3}{4\pi B \bar{b}}\left(\d_z^2 \dot{u}_2+\d_{\bar z}^2 \dot{u}_{-2}\right).
\end{equation}
We leave the detail derivation of above equation to Appendix \ref{sec:density}. Using the explicit value of $k_F$ from Eq. \eqref{eq:kF}, we see that only term including $u_{\pm 2}$ in \eqref{eq:rhodot} survives and we are left with
\begin{equation}
\delta \rho^{(1)}=\frac{k_F^3}{4\pi B \bar{b}}\left(\d_z^2  u_2+\d_{\bar z}^2  u_{-2}\right).
\end{equation} 
Note that by the construction of the Hamiltonian \eqref{eq:Ham}, at
long wave length limit the ground state is annihilated by $u_m$ with
$m<0$ (see Ref.~\cite{Nguyen2018b} for details):
$u_{-2}|0\>=\<0|u_2=0$, which allows us to derive the equal-time
density-density correlation function,
\begin{equation}
\<\delta \rho^{(1)}(-\mathbf{q})\delta \rho^{(1)}(\mathbf{q}) \>=q^4\frac{k_F^6}{256 \pi^2 B^2 \bar{b}^2}\< [u_{-2}(-\mathbf{q}),u_2(\mq)]\>.
\end{equation}
We then use the commutation relation of $u_{\pm 2}$ and obtain
	\begin{equation}
		\label{eq:density-density}
\<\delta \rho^{(1)}(-\mathbf{q})\delta \rho^{(1)}(\mathbf{q}) \>=\frac{k_F^4}{64\pi B^2|\bar{b}|}q^4=\frac{\pi}{4}\frac{\bar{\rho}^2_{\text{\tiny{CF}}}}{|\bar{b}|}(q \ell_B)^4,
	\end{equation}
From Eq.~\eqref{eq:density-density} we then read off the contribution
to the coefficient of $q^4 \ell_B^4$ of the projected SSF from the CF
sector
	\begin{equation}
		\label{eq:CFs4}
	\bar{s}_4^{(1)}=\frac{1}{8}\frac{\bar{\rho}^2_{\text{\tiny{CF}}}}{\bar{\rho}_e}\frac{2\pi }{|\bar{b}|},
	\end{equation}
where we defined the projected SSF following Girvin, MacDonald, and
Platzman~\cite{GMP:1986}
\begin{equation}
\bar{s}=\frac{1}{\bar{\rho}_e}\<\delta \rho(-\mathbf{q})\delta \rho(\mathbf{q}) \>.
\end{equation}

From the action of the Haldane sector \eqref{eq:Haldane}, one can
calculate the contribution to the equal time correlation function
$\<\delta \rho^{(2)}(-\mathbf{q})\delta \rho^{(2)}(\mathbf{q}) \>$ as
well as the contribution of the Haldane sector to the projected
SSF. The calculation follows directly from Ref. \cite{Gromov2017}, we
will not repeat the calculation and quote the result only:
\begin{equation}
	\label{eq:Haldanes4}
  \bar{s}_4^{(2)}=\frac{2 \zeta}{16n \nu_+} .
\end{equation}
We then have the total static structure factor
\footnote{
The mixed correlation $\<\rho^{(1)}(-\mq)\rho^{(2)}(\mq) \>$ vanishes.}
	\begin{equation}
		\label{eq:s4full}
	\bar{s}_4=	\bar{s}_4^{(1)}+\bar{s}_4^{(2)}=\frac{1}{8}\left(\frac{\bar{\rho}^2_{\text{\tiny{CF}}}}{\bar{\rho}_e}\frac{2\pi }{|\bar{b}|}+\frac{2 \zeta}{2n \nu_+}\right). 
	\end{equation}
We are now in position to compute the coefficient $\bar{s}_4$ for both
filling fraction $\nu_{\pm}$ of the general Jain's
sequences. Combining Eqs.~\eqref{eq:s4full}, \eqref{eq:rhoCF},
\eqref{eq:rhoe} and \eqref{eq:bbar} one obtain, for filling fraction
$\nu_+=\frac{N}{2nN+1}$
\begin{equation}\label{eq:s4pN}
  \bar{s}_4^{\nu_+}=\frac{1}{8}\left(N+1+2\zeta
  \right) + O\left( \frac1N\right),
\end{equation}
and for $\nu_-=\frac{N}{2nN-1}$,
\begin{equation}\label{eq:s4mN}
  \bar{s}_4^{\nu_-}=\frac{1}{8}\left(N-1+2\zeta
  \right) + O\left( \frac1N\right).
\end{equation}
The $O(1/N)$ corrections in Eqs.~(\ref{eq:s4pN}) and (\ref{eq:s4mN})
are beyond the limit of accuracy of the semiclassical approximation
and will be dropped in future formulas.  If we fix the coefficient
$\zeta=n-1$ as in Section \ref{sec:WZ} to match the Wen-Zee shift, we
finally have the coefficient $\bar{s}_4$,
\begin{align}
	\bar{s}_4^{\nu_+}& = \frac{1}{8} (N+2n-1),\\
	\bar{s}_4^{\nu_-}& = \frac{1}{8} (N-3+2n).
\end{align}
Remembering that the shift of $\nu_+$ is $\cS_{\nu_+}=N+2n$, we see that $\bar{s}^{\nu_+}_4$ saturates the Haldane bound
\begin{equation}
	\bar{s}^{\nu_+}_4 = \frac{\cS_{\nu_+}-1}{8},
\end{equation}
which suggests that $\nu_+$ is a chiral state \cite{Haldane:2009idea,Nguyen:2014LLL}. However, the Haldane bound for the $\nu_-$ state,
\begin{equation}
	\bar{s}^{\nu_-}_4\ge \frac{|\cS_{\nu_-}-1|}{8},
\end{equation}
is saturated only for $n=1$ \cite{Nguyen:PH}.  For example, the $\nu=N/(4N-1)$ states are
not chiral: the chirality of the Haldane mode is opposite to the chirality
of the low-energy magnetoroton.
	
We emphasize here that the contribution to projected SSF of the
Haldane sector is crucial in our proposal.  Without the additional
term $\bar{s}^{(2)}_4$, the projected SSF of the Jain sequence $\nu_+$
violates the Haldane bound for $n>1$
\cite{Haldane:2009idea,Haldane2011}. The coefficient $\zeta=n-1$ not
only helps us to obtain the correct Wen-Zee shift in Sec. \ref{sec:WZ}
but also makes it possible to satisfy the lower bound on the projected SSF.
	
\section{Conclusion}
\label{sec:concl}

In this paper, we propose a Dirac composite fermion model of general
Jain's sequences $\nu=\frac{N}{2n N \pm 1}$.
The proposal incorporates a Dirac composite fermion and an extra spin-2 mode at $\nu=1/4$ and the Jain sequences around it.  We show that our model reproduces the known results of Hall conductivity and Hall
viscosity. The results imply that our model satisfies the Galilean
invariant and reproduces the Wen-Zee shift.  Using the semiclassical
approach, we calculated the coefficient $\bar{s}_4$ of the projected
static structure factor.  We show that, without the extra mode, the
projected SSF violates the Haldane bound.  We also derive the
long-wavelength limit of the GMP algebra that emerges from the model. In summary, the results calculated in this paper suggest that our Dirac
composite fermion model is a proper effective theory of LLL that
provides the expected physical results of general Jain's sequences.

In this paper, we proposed a single Haldane mode with spin-2. However, we can come up with the Haldane sector that includes multiple extra spin-2 modes. In particular, if we consider $s$ spin-2 modes with different energy gaps determined by $\tilde{m}_i$ and different coupling $\zeta_i$ with background electromagnetic field as described in Eq. \eqref{eq:Haldane}, then we can reproduce the physical results of this paper with the condition 
\begin{equation}
	\sum_{i=-1}^s \zeta_i = n-1.
\end{equation} 
We then arrive at the other main conclusion of this paper, which is the extra spin-2 mode(s). We expect that those modes have
higher energy gaps than the Dirac composite fermion
effective cyclotron frequency but still lower than the electron's
cyclotron gap. We suggest that the appearance of the extra spin-2
mode(s) can be confirmed in the numerical calculation of the correlation function of the stress tensor operator in the LLL
\cite{Liou2019,haldane2021} at or near the filling fraction $\frac{1}{4}$.
We expect that there would be an additional peak(s) in the spectral
function of the stress tensor at high frequency. We can also verify
the extra spin-2 mode(s) with their chiralities by circularly
polarized Raman scattering \cite{Raman}.

In this paper, we assume that the Berry phase of the composite fermion is equal to $\pi$ as for a massless Dirac fermion, a choice motivated by
the symmetry of the $I-V$ curve around $\nu=1/4$.
Without this motivation, one can consider other possibilities; for example, 
one in which the CF has zero Berry phase as in the HLR theory.  
The Haldane mode will also be needed here to avoid violation of the
Haldane bound within the HLR theory (which already occurs at near half
filling~\cite{Nguyen2018a}).


\acknowledgments

We thank Andrey Gromov, Duncan Haldane, Ed Rezayi, and Kun Yang for discussions.  DTS
is supported in part, by the U.S.\ DOE grant No.\ DE-FG02-13ER41958, a
Simons Investigator grant and by the Simons Collaboration on
Ultra-Quantum Matter, which is a grant from the Simons Foundation
(651440, DTS).  DXN was supported by Brown Theoretical Physics Center.

\appendix
\section{Spectral sum rules and the Haldane bound}
\label{sec:HB}
For completeness, we present the derivation of the Haldane bound in this Appendix. 2D electrons in a constant magnetic field satisfies the following Ward's identities 
\begin{align}
	\partial_t \rho+\mathbf{\nabla}\cdot \mathbf{j}&=0 \\
	\partial_t (mj_i) + \partial_k T_{ki} &= (\mathbf{j}\times \mathbf B)_i
\end{align}
which are nothing but the conservations of $U(1)$ charge and momentum density, where the momentum density of non-relativistic electron is  $P_i=m j_i$. In the lowest Landau level limit, where we take electron's mass to zero, the momentum density conservation equation becomes the force balance equation from which one can derive $j_i$ in terms of $T_{ij}$. Combining with the charge conservation, one obtain the lowest Landau level Ward's identity \cite{Nguyen:2014LLL}
\begin{equation}
	\label{eq:LLWard}
	\partial_t \rho =\frac{4i}{B}\left(\partial_{\bar z} \partial_{\bar z} T_{z z}-\partial_z \partial_z T_{\bar z \bar z}\right)
\end{equation} 
Using the Ward identity \eqref{eq:LLWard}, one can derive the first spectral sum rule for lowest Landau level \cite{Nguyen:2014LLL,Golkar2016} \footnote{The upper limit $\infty$ here means a frequency which is much higher than the Coulomb gap but also smaller the the cyclotron gap. }
\begin{equation}
	\label{eq:sr1}
	\int_0^\infty \frac{d\omega}{\omega^2} \left[
	\rho_T(\omega) + \bar\rho_T(\omega) \right] = s_4,
\end{equation}
with $s_4$ is the coefficient of $q^4 \ell_B^4$ of the projected SSF and the spectral functions $\rho_T$ and $\bar{\rho}_T$ are defined as \cite{Golkar2016}
\begin{align}
	\rho_T(\omega) &= \frac1N \sum_n |\< n| \int\!d\x\, T_{zz} |0\>|^2 \delta(\omega-E_n),\\
	\bar\rho_T(\omega) &= \frac1N \sum_n |\< n| \int\!d\x\, T_{\bar z\bar z} |0\>|^2 \delta(\omega-E_n),
\end{align}
where $N$ is the total number of electrons, $|0\>$ is the ground
state, the sum is taken over all excited states $|n\>$ in the lowest Landau level, and $E_n$ is
the energy of the state $|n\>$. The relation of Hall viscosity in term of the retard Green's functions of stress tensor at zero spatial momentum 
\begin{equation}
	\label{eq:Hallvis1}
	\omega \eta_H(\omega)=\langle T_{zz} T_{{\bar z} {\bar z}}\rangle_\omega-\langle T_{{\bar z} {\bar z}}T_{zz} \rangle_\omega
\end{equation}
with the definition of retard Green's function at frequency $\omega$
\begin{equation}
	\langle A B\rangle_\omega=-i \int \, d t d^2 \,\mathbf{x}  e^{i\omega t} \Theta(t) \langle A(t,\mathbf{x}) B(0,\mathbf{0}) \rangle
\end{equation}
Using Eq \eqref{eq:Hallvis1}, one can derive the second lowest Landau level sum rule
\begin{equation}
	\label{eq:sr2}
	\int_0^\infty \frac{d\omega}{\omega^2} \left[
	\rho_T(\omega) - \bar\rho_T(\omega) \right] =\frac{\eta_H(0)-\eta_H(\infty)}{2\bar{\rho}_e}=\frac{\mathcal{S}-1}{8},
\end{equation}
where we used the value of Hall viscosity at zero and high frequency
\begin{equation}
	\eta_H(0)=\bar{\rho}_e\frac{\mathcal{S}}{4}, \quad \eta_H(\infty)=\frac{\bar{\rho}_e}{4}.
\end{equation}
By definition, both $\rho_T(\omega)$ and $\bar\rho_T(\omega)$ are non-negative, therefore the combination of sum rules \eqref{eq:sr1}, \eqref{eq:sr2} leads us to an inequality 
\begin{equation}
	s_4 \geq \frac{|\mathcal{S}-1|}{8},
\end{equation} 
which is the Haldane bound.

\section{GMP algebra of the Dirac composite fermion sector}
\label{sec:GMPCF}

In this Appendix, we provide a detailed derivation of the GMP algebra
of the composite fermion sector [Eq.~\eqref{eq:GMPCF} in the main
text]. We ignore the trivial contribution of $B$ to the charge density
operator of CF section and rewrite
\begin{equation}
	\label{eq:rho1}
\rho^{(1)}=-\epsilon_{ij} \d_i \tilde{\Pi}_j ,
\end{equation} 
with
\begin{align}
\tilde{\Pi}_i & =\frac{1}{B}\Pi_i +\frac{1}{4\pi n} a_i =\frac{1}{B}\left[\tilde{\pi}_i - \left(\rho_{\text{\tiny{CF}}}{-}\frac{B}{4\pi n}\right)a_i\right] ,
\end{align}
where we have used $\tilde{\pi}_i=-\frac{i}{2}\psi^\dagger
\overleftrightarrow{\d}_{\!\!i} \psi$. Using Eqs.~\eqref{eq:commutepsi} and
\eqref{eq:commuteai}, one obtains the following commutation relations
by direct calculations
\begin{multline}
	\label{eq:compi}
	[\tilde\pi_i(\mx), \, \tilde\pi_j(\my) ] =\\
	- i \left [ \tilde\pi_j(\mx) \frac\d{\d x^i}
	-\tilde\pi_i(\my) \frac\d{\d y^j} \right]\delta(\mx-\my), 
\end{multline}
\begin{multline}
	\label{eq:compia}
	\left[\tilde{\pi}_i(\mx), \, -\Bigl(\rho_{\text{\tiny{CF}}}(\my)-\frac{B}{4\pi n}\Bigr)a_j(\my)\right]=\\i \rho_{\text{\tiny{CF}}}(\mx) a_j(\mx) \frac{\d}{\d x^i}\delta(\mx-\my)
	+ i \rho_{\text{\tiny{CF}}}(\mx) \d_i a_j(\mx) \delta(\mx-\my),
\end{multline}
\begin{multline}
	\label{eq:comapi}
	\left[-\Bigl(\rho_{\text{\tiny{CF}}}(\mx)-\frac{B}{4\pi n}\Bigr)a_i(\mx),\, \tilde{\pi}_j(\my)\right]=\\-i \rho_{\text{\tiny{CF}}}(\my) a_i(\my) \frac{\d}{\d y^j}\delta(\mx-\my)
	- i \rho_{\text{\tiny{CF}}}(\my) \d_j a_i(\my) \delta(\mx-\my),
\end{multline}
\begin{align}
	\label{eq:comaa}
	&\left[-(\rho_{\text{\tiny{CF}}}(\mx)-\frac{B}{4\pi n})a_i(\mx),\, -(\rho_{\text{\tiny{CF}}}(\my)-\frac{B}{4\pi n})a_j(\my)\right]\nonumber\\&=-i \frac{4\pi n}{n-1}\epsilon^{ij}\delta(\x-\y)
        \Bigl(\rho_{\text{\tiny{CF}}}(\mx)-\frac{B}{4\pi n}\Bigr)^2 \nonumber\\ &= -i b(\mx)\epsilon^{ij}\delta(\x-\y)\Bigl(\rho_{\text{\tiny{CF}}}(\mx)-\frac{B}{4\pi n}\Bigr),
\end{align}
where we have used the constraint 
\begin{equation}
	\label{eq:rhoCF1}
\rho_{\text{\tiny{CF}}}(\mx)=\frac{n-1}{4\pi n} b(\mx)+\frac{B}{4\pi n}
\end{equation}
to arrive to the last form in Eq.~\eqref{eq:comaa}. Combining
Eqs.~\eqref{eq:compi}, \eqref{eq:compia}, \eqref{eq:comapi},
\eqref{eq:comaa} and the identity
\begin{equation}
  f(\x) \frac\d{\d x^i}\delta(\x-\y) = f(\y) \frac\d{\d x^i} \delta(\x-\y)
  - \d_i f (\mx) \delta(\x-\y),
\end{equation}
we obtain the commutation relation
\begin{multline}
	\label{eq:comPi}
	[\tilde{\Pi}_i(\x), \, \tilde{\Pi}_j(\y)] =
	-\frac{i}{B} \left[ \Pi_j(\x)\frac\d{\d x^i} - \Pi_i(\y) \frac\d{\d y^j}\right]
	\delta(\x-\y)\\
	+i \frac{B}{4\pi n}\left[ a_j(\y)\frac\d{\d x^i} 
	-  a_i(\x) \frac\d{\d y^j} \right] \delta(\x-\y).
\end{multline}
 Combining \eqref{eq:comPi} and \eqref{eq:rho1}, we obtain the commutation relation of charge density in the momentum space \footnote{The extra terms (second line in the equation \eqref{eq:comPi}) disappear when we
 	compute the curl's in $[\rho^{(1)}(\x), \, \rho^{(1)}(\y)]$.}
\begin{align}
	\label{eq:comrhorho}
[\rho^{(1)}(\mk),\rho^{(1)}(\mq)] =& i (\mk \times \mq)\ell^2_B\rho^{(1)}(\mk+\mq).
\end{align}
which is Eq.~\eqref{eq:GMPCF} of the main text. 
\section{Density operator in the semiclassical formalism}
\label{sec:density}
In this Appendix, we provide a detailed derivation of the charge
density operator of the CF sector in terms of $u_m$ operators. From
Eqs.~\eqref{eq:rhoCF1} and \eqref{eq:semiope} we have
\begin{equation}
	\label{eq:u0new}
	\dot{u}_0=\frac{n-1}{2 n k_F}\dot{b}.
\end{equation}
We rewrite the equation of motion for $u_{\pm 1}$ as
\begin{multline}
	\label{eq:u1}
	\dot{u}_1=-i(1+F_1)\frac{\bar{ b} v_F}{k_F}u_1-v_F(1+F_0)\d_{\bar z}u_0\\-v_F(1+F_2)\d_z u_2 +e_{\bar z},
\end{multline}
\begin{multline}
		\label{eq:um1}
	\dot{u}_{-1}=i(1+F_1)\frac{\bar{ b} v_F}{k_F}u_{-1}-v_F(1+F_0)\d_{ z}u_0\\-v_F(1+F_2)\d_{\bar z} u_2 +e_{\bar z}.
\end{multline}
We also need the equation of motion for $u_{\pm 2}$ up to leading
order in spatial derivatives
\begin{align}
	\label{eq:u2}
\dot{u}_2&=-i(1+F_2)\frac{\bar{ b} v_F}{k_F} u_2 ,\\
	\label{eq:um2}
\dot{u}_{-2}&=i(1+F_2)\frac{\bar{ b} v_F}{k_F} u_2 .
\end{align}
Combing the time derivative of Eq.~\eqref{eq:semi-rho} and Eqs. \eqref{eq:u1}, \eqref{eq:um1}, we obtain 
\begin{multline}
\dot{\rho}^{(1)}=-\frac{\dot{b}}{4\pi n}+\frac{k_F v_F \bar{b}}{2 \pi B}(1+F_1)(\d_z u_1 + \d_{\bar z}u_{-1})\\
-\frac{i v_F k_F^2}{2 \pi B}(1+F_2)(\d_z^2 u_2 - \d_{\bar z}u_{-2})+\frac{i k_F^2}{2 \pi B}(\d_z e_{\bar z}-\d_{\bar z}e_z).
\end{multline}
With the help of the Bianchi identity $\d_z e_{\bar z}-\d_{\bar
  z}e_z=-\frac{i}{2}\dot{b}$, the equation of motion for $u_{\pm 2}$,
and a combination of Eqs.~\eqref{eq:u0} and \eqref{eq:u0new}, we
obtain
\begin{equation}
	\dot{\rho}^{(1)}=\frac{\dot{b}}{4 \pi}\left(\frac{k_F^2}{B}-\frac{1}{n}-\frac{\bar b}{B}\frac{n-1}{n}\right)+\frac{k_F^3}{4\pi B \bar{b}}\left(\d_z^2 \dot{u}_2+\d_{\bar z}^2 \dot{u}_{-2}\right),
\end{equation}
  which is Eq.~\eqref{eq:rhodot} in the main text. 
  
\bibliography{DCFeven}

\begin{thebibliography}{56}%
\makeatletter
\providecommand \@ifxundefined [1]{%
 \@ifx{#1\undefined}
}%
\providecommand \@ifnum [1]{%
 \ifnum #1\expandafter \@firstoftwo
 \else \expandafter \@secondoftwo
 \fi
}%
\providecommand \@ifx [1]{%
 \ifx #1\expandafter \@firstoftwo
 \else \expandafter \@secondoftwo
 \fi
}%
\providecommand \natexlab [1]{#1}%
\providecommand \enquote  [1]{``#1''}%
\providecommand \bibnamefont  [1]{#1}%
\providecommand \bibfnamefont [1]{#1}%
\providecommand \citenamefont [1]{#1}%
\providecommand \href@noop [0]{\@secondoftwo}%
\providecommand \href [0]{\begingroup \@sanitize@url \@href}%
\providecommand \@href[1]{\@@startlink{#1}\@@href}%
\providecommand \@@href[1]{\endgroup#1\@@endlink}%
\providecommand \@sanitize@url [0]{\catcode `\\12\catcode `\$12\catcode
  `\&12\catcode `\#12\catcode `\^12\catcode `\_12\catcode `\%12\relax}%
\providecommand \@@startlink[1]{}%
\providecommand \@@endlink[0]{}%
\providecommand \url  [0]{\begingroup\@sanitize@url \@url }%
\providecommand \@url [1]{\endgroup\@href {#1}{\urlprefix }}%
\providecommand \urlprefix  [0]{URL }%
\providecommand \Eprint [0]{\href }%
\providecommand \doibase [0]{https://doi.org/}%
\providecommand \selectlanguage [0]{\@gobble}%
\providecommand \bibinfo  [0]{\@secondoftwo}%
\providecommand \bibfield  [0]{\@secondoftwo}%
\providecommand \translation [1]{[#1]}%
\providecommand \BibitemOpen [0]{}%
\providecommand \bibitemStop [0]{}%
\providecommand \bibitemNoStop [0]{.\EOS\space}%
\providecommand \EOS [0]{\spacefactor3000\relax}%
\providecommand \BibitemShut  [1]{\csname bibitem#1\endcsname}%
\let\auto@bib@innerbib\@empty
\bibitem [{\citenamefont {Tsui}\ \emph {et~al.}(1982)\citenamefont {Tsui},
  \citenamefont {Stormer},\ and\ \citenamefont {Gossard}}]{FQH1}%
  \BibitemOpen
  \bibfield  {author} {\bibinfo {author} {\bibfnamefont {D.~C.}\ \bibnamefont
  {Tsui}}, \bibinfo {author} {\bibfnamefont {H.~L.}\ \bibnamefont {Stormer}},\
  and\ \bibinfo {author} {\bibfnamefont {A.~C.}\ \bibnamefont {Gossard}},\
  }\bibfield  {title} {\bibinfo {title} {{Two-Dimensional Magnetotransport in
  the Extreme Quantum Limit}},\ }\href
  {https://doi.org/10.1103/PhysRevLett.48.1559} {\bibfield  {journal} {\bibinfo
   {journal} {Phys. Rev. Lett.}\ }\textbf {\bibinfo {volume} {48}},\ \bibinfo
  {pages} {1559} (\bibinfo {year} {1982})}\BibitemShut {NoStop}%
\bibitem [{\citenamefont {Laughlin}(1983)}]{FQH2}%
  \BibitemOpen
  \bibfield  {author} {\bibinfo {author} {\bibfnamefont {R.~B.}\ \bibnamefont
  {Laughlin}},\ }\bibfield  {title} {\bibinfo {title} {{Anomalous Quantum Hall
  Effect: An Incompressible Quantum Fluid with Fractionally Charged
  Excitations}},\ }\href {https://doi.org/10.1103/PhysRevLett.50.1395}
  {\bibfield  {journal} {\bibinfo  {journal} {Phys. Rev. Lett.}\ }\textbf
  {\bibinfo {volume} {50}},\ \bibinfo {pages} {1395} (\bibinfo {year}
  {1983})}\BibitemShut {NoStop}%
\bibitem [{\citenamefont {Jain}(1989)}]{Jain1}%
  \BibitemOpen
  \bibfield  {author} {\bibinfo {author} {\bibfnamefont {J.~K.}\ \bibnamefont
  {Jain}},\ }\bibfield  {title} {\bibinfo {title} {{Composite-fermion approach
  for the fractional quantum Hall effect}},\ }\href
  {https://doi.org/10.1103/PhysRevLett.63.199} {\bibfield  {journal} {\bibinfo
  {journal} {Phys. Rev. Lett.}\ }\textbf {\bibinfo {volume} {63}},\ \bibinfo
  {pages} {199} (\bibinfo {year} {1989})}\BibitemShut {NoStop}%
\bibitem [{\citenamefont {Wilczek}(1982)}]{Flux1}%
  \BibitemOpen
  \bibfield  {author} {\bibinfo {author} {\bibfnamefont {F.}~\bibnamefont
  {Wilczek}},\ }\bibfield  {title} {\bibinfo {title} {{Magnetic Flux, Angular
  Momentum, and Statistics}},\ }\href
  {https://doi.org/10.1103/PhysRevLett.48.1144} {\bibfield  {journal} {\bibinfo
   {journal} {Phys. Rev. Lett.}\ }\textbf {\bibinfo {volume} {48}},\ \bibinfo
  {pages} {1144} (\bibinfo {year} {1982})}\BibitemShut {NoStop}%
\bibitem [{\citenamefont {Halperin}(1983)}]{Flux2}%
  \BibitemOpen
  \bibfield  {author} {\bibinfo {author} {\bibfnamefont {B.}~\bibnamefont
  {Halperin}},\ }\bibfield  {title} {\bibinfo {title} {{Theory of the quantized
  Hall conductance}},\ }\href
  {https://www.e-periodica.ch/digbib/view?pid=hpa-001:1983:56::1243} {\bibfield
   {journal} {\bibinfo  {journal} {Helv. Phys. Acta}\ }\textbf {\bibinfo
  {volume} {56}},\ \bibinfo {pages} {75} (\bibinfo {year} {1983})}\BibitemShut
  {NoStop}%
\bibitem [{\citenamefont {Girvin}\ and\ \citenamefont
  {MacDonald}(1987)}]{Flux3}%
  \BibitemOpen
  \bibfield  {author} {\bibinfo {author} {\bibfnamefont {S.~M.}\ \bibnamefont
  {Girvin}}\ and\ \bibinfo {author} {\bibfnamefont {A.~H.}\ \bibnamefont
  {MacDonald}},\ }\bibfield  {title} {\bibinfo {title} {{Off-diagonal
  long-range order, oblique confinement, and the fractional quantum Hall
  effect}},\ }\href {https://doi.org/10.1103/PhysRevLett.58.1252} {\bibfield
  {journal} {\bibinfo  {journal} {Phys. Rev. Lett.}\ }\textbf {\bibinfo
  {volume} {58}},\ \bibinfo {pages} {1252} (\bibinfo {year}
  {1987})}\BibitemShut {NoStop}%
\bibitem [{\citenamefont {Halperin}\ \emph {et~al.}(1993)\citenamefont
  {Halperin}, \citenamefont {Lee},\ and\ \citenamefont {Read}}]{HLR}%
  \BibitemOpen
  \bibfield  {author} {\bibinfo {author} {\bibfnamefont {B.~I.}\ \bibnamefont
  {Halperin}}, \bibinfo {author} {\bibfnamefont {P.~A.}\ \bibnamefont {Lee}},\
  and\ \bibinfo {author} {\bibfnamefont {N.}~\bibnamefont {Read}},\ }\bibfield
  {title} {\bibinfo {title} {{Theory of the half-filled Landau level}},\ }\href
  {https://doi.org/10.1103/PhysRevB.47.7312} {\bibfield  {journal} {\bibinfo
  {journal} {Phys. Rev. B}\ }\textbf {\bibinfo {volume} {47}},\ \bibinfo
  {pages} {7312} (\bibinfo {year} {1993})}\BibitemShut {NoStop}%
\bibitem [{\citenamefont {Kang}\ \emph {et~al.}(1993)\citenamefont {Kang},
  \citenamefont {Stormer}, \citenamefont {Pfeiffer}, \citenamefont {Baldwin},\
  and\ \citenamefont {West}}]{Kang}%
  \BibitemOpen
  \bibfield  {author} {\bibinfo {author} {\bibfnamefont {W.}~\bibnamefont
  {Kang}}, \bibinfo {author} {\bibfnamefont {H.~L.}\ \bibnamefont {Stormer}},
  \bibinfo {author} {\bibfnamefont {L.~N.}\ \bibnamefont {Pfeiffer}}, \bibinfo
  {author} {\bibfnamefont {K.~W.}\ \bibnamefont {Baldwin}},\ and\ \bibinfo
  {author} {\bibfnamefont {K.~W.}\ \bibnamefont {West}},\ }\bibfield  {title}
  {\bibinfo {title} {{How real are composite fermions?}},\ }\href
  {https://doi.org/10.1103/PhysRevLett.71.3850} {\bibfield  {journal} {\bibinfo
   {journal} {Phys. Rev. Lett.}\ }\textbf {\bibinfo {volume} {71}},\ \bibinfo
  {pages} {3850} (\bibinfo {year} {1993})}\BibitemShut {NoStop}%
\bibitem [{\citenamefont {Kivelson}\ \emph {et~al.}(1997)\citenamefont
  {Kivelson}, \citenamefont {Lee}, \citenamefont {Krotov},\ and\ \citenamefont
  {Gan}}]{PHKivelson}%
  \BibitemOpen
  \bibfield  {author} {\bibinfo {author} {\bibfnamefont {S.~A.}\ \bibnamefont
  {Kivelson}}, \bibinfo {author} {\bibfnamefont {D.-H.}\ \bibnamefont {Lee}},
  \bibinfo {author} {\bibfnamefont {Y.}~\bibnamefont {Krotov}},\ and\ \bibinfo
  {author} {\bibfnamefont {J.}~\bibnamefont {Gan}},\ }\bibfield  {title}
  {\bibinfo {title} {{Composite-fermion Hall conductance at $\nu=\tfrac12$}},\
  }\href {https://doi.org/10.1103/PhysRevB.55.15552} {\bibfield  {journal}
  {\bibinfo  {journal} {Phys. Rev. B}\ }\textbf {\bibinfo {volume} {55}},\
  \bibinfo {pages} {15552} (\bibinfo {year} {1997})}\BibitemShut {NoStop}%
\bibitem [{\citenamefont {Kamburov}\ \emph {et~al.}(2014)\citenamefont
  {Kamburov}, \citenamefont {Liu}, \citenamefont {Mueed}, \citenamefont
  {Shayegan}, \citenamefont {Pfeiffer}, \citenamefont {West},\ and\
  \citenamefont {Baldwin}}]{Reflection3}%
  \BibitemOpen
  \bibfield  {author} {\bibinfo {author} {\bibfnamefont {D.}~\bibnamefont
  {Kamburov}}, \bibinfo {author} {\bibfnamefont {Y.}~\bibnamefont {Liu}},
  \bibinfo {author} {\bibfnamefont {M.~A.}\ \bibnamefont {Mueed}}, \bibinfo
  {author} {\bibfnamefont {M.}~\bibnamefont {Shayegan}}, \bibinfo {author}
  {\bibfnamefont {L.~N.}\ \bibnamefont {Pfeiffer}}, \bibinfo {author}
  {\bibfnamefont {K.~W.}\ \bibnamefont {West}},\ and\ \bibinfo {author}
  {\bibfnamefont {K.~W.}\ \bibnamefont {Baldwin}},\ }\bibfield  {title}
  {\bibinfo {title} {{What Determines the Fermi Wave Vector of Composite
  Fermions?}},\ }\href {https://doi.org/10.1103/PhysRevLett.113.196801}
  {\bibfield  {journal} {\bibinfo  {journal} {Phys. Rev. Lett.}\ }\textbf
  {\bibinfo {volume} {113}},\ \bibinfo {pages} {196801} (\bibinfo {year}
  {2014})}\BibitemShut {NoStop}%
\bibitem [{\citenamefont {Son}(2015)}]{Son:Dirac}%
  \BibitemOpen
  \bibfield  {author} {\bibinfo {author} {\bibfnamefont {D.~T.}\ \bibnamefont
  {Son}},\ }\bibfield  {title} {\bibinfo {title} {{Is the Composite Fermion a
  Dirac Particle?}},\ }\href {https://doi.org/10.1103/PhysRevX.5.031027}
  {\bibfield  {journal} {\bibinfo  {journal} {Phys. Rev. X}\ }\textbf {\bibinfo
  {volume} {5}},\ \bibinfo {pages} {031027} (\bibinfo {year}
  {2015})}\BibitemShut {NoStop}%
\bibitem [{\citenamefont {Nguyen}\ \emph
  {et~al.}(2018{\natexlab{a}})\citenamefont {Nguyen}, \citenamefont {Golkar},
  \citenamefont {Roberts},\ and\ \citenamefont {Son}}]{Nguyen2018a}%
  \BibitemOpen
  \bibfield  {author} {\bibinfo {author} {\bibfnamefont {D.~X.}\ \bibnamefont
  {Nguyen}}, \bibinfo {author} {\bibfnamefont {S.}~\bibnamefont {Golkar}},
  \bibinfo {author} {\bibfnamefont {M.~M.}\ \bibnamefont {Roberts}},\ and\
  \bibinfo {author} {\bibfnamefont {D.~T.}\ \bibnamefont {Son}},\ }\bibfield
  {title} {\bibinfo {title} {{Particle-hole symmetry and composite fermions in
  fractional quantum Hall states}},\ }\href
  {https://doi.org/10.1103/physrevb.97.195314} {\bibfield  {journal} {\bibinfo
  {journal} {Phys. Rev. B}\ }\textbf {\bibinfo {volume} {97}},\ \bibinfo
  {pages} {195314} (\bibinfo {year} {2018}{\natexlab{a}})}\BibitemShut
  {NoStop}%
\bibitem [{\citenamefont {Shahar}\ \emph {et~al.}(1996)\citenamefont {Shahar},
  \citenamefont {Tsui}, \citenamefont {Shayegan}, \citenamefont {Shimshoni},\
  and\ \citenamefont {Sondhi}}]{Reflection2}%
  \BibitemOpen
  \bibfield  {author} {\bibinfo {author} {\bibfnamefont {D.}~\bibnamefont
  {Shahar}}, \bibinfo {author} {\bibfnamefont {D.~C.}\ \bibnamefont {Tsui}},
  \bibinfo {author} {\bibfnamefont {M.}~\bibnamefont {Shayegan}}, \bibinfo
  {author} {\bibfnamefont {E.}~\bibnamefont {Shimshoni}},\ and\ \bibinfo
  {author} {\bibfnamefont {S.~L.}\ \bibnamefont {Sondhi}},\ }\bibfield  {title}
  {\bibinfo {title} {{Evidence for Charge-Flux Duality near the Quantum Hall
  Liquid-to-Insulator Transition}},\ }\href
  {https://doi.org/10.1126/science.274.5287.589} {\bibfield  {journal}
  {\bibinfo  {journal} {Science}\ }\textbf {\bibinfo {volume} {274}},\ \bibinfo
  {pages} {589} (\bibinfo {year} {1996})}\BibitemShut {NoStop}%
\bibitem [{\citenamefont {Wang}(2019)}]{Wang:2019}%
  \BibitemOpen
  \bibfield  {author} {\bibinfo {author} {\bibfnamefont {J.}~\bibnamefont
  {Wang}},\ }\bibfield  {title} {\bibinfo {title} {{Dirac Fermion Hierarchy of
  Composite Fermi Liquids}},\ }\href
  {https://doi.org/10.1103/PhysRevLett.122.257203} {\bibfield  {journal}
  {\bibinfo  {journal} {Phys. Rev. Lett.}\ }\textbf {\bibinfo {volume} {122}},\
  \bibinfo {pages} {257203} (\bibinfo {year} {2019})}\BibitemShut {NoStop}%
\bibitem [{Note1()}]{Note1}%
  \BibitemOpen
  \bibinfo {note} {We will leave an explicit derivation of the Haldane bound to
  Appendix \ref {sec:HB}}\BibitemShut {NoStop}%
\bibitem [{\citenamefont {Goldman}\ and\ \citenamefont
  {Fradkin}(2018)}]{Goldman2018}%
  \BibitemOpen
  \bibfield  {author} {\bibinfo {author} {\bibfnamefont {H.}~\bibnamefont
  {Goldman}}\ and\ \bibinfo {author} {\bibfnamefont {E.}~\bibnamefont
  {Fradkin}},\ }\bibfield  {title} {\bibinfo {title} {{Dirac composite fermions
  and emergent reflection symmetry about even-denominator filling fractions}},\
  }\href {https://doi.org/10.1103/physrevb.98.165137} {\bibfield  {journal}
  {\bibinfo  {journal} {Phys. Rev. B}\ }\textbf {\bibinfo {volume} {98}},\
  \bibinfo {pages} {165137} (\bibinfo {year} {2018})}\BibitemShut {NoStop}%
\bibitem [{\citenamefont {Haldane}(2009)}]{Haldane:2009idea}%
  \BibitemOpen
  \bibfield  {author} {\bibinfo {author} {\bibfnamefont {F.~D.~M.}\
  \bibnamefont {Haldane}},\ }\href@noop {} {\bibinfo {title} {{``Hall
  viscosity'' and intrinsic metric of incompressible fractional Hall fluids}}}
  (\bibinfo {year} {2009}),\ \Eprint {https://arxiv.org/abs/0906.1854}
  {arXiv:0906.1854} \BibitemShut {NoStop}%
\bibitem [{\citenamefont {Gromov}\ and\ \citenamefont
  {Son}(2017)}]{Gromov2017}%
  \BibitemOpen
  \bibfield  {author} {\bibinfo {author} {\bibfnamefont {A.}~\bibnamefont
  {Gromov}}\ and\ \bibinfo {author} {\bibfnamefont {D.~T.}\ \bibnamefont
  {Son}},\ }\bibfield  {title} {\bibinfo {title} {{Bimetric Theory of
  Fractional Quantum Hall States}},\ }\href
  {https://doi.org/10.1103/physrevx.7.041032} {\bibfield  {journal} {\bibinfo
  {journal} {Phys. Rev. X}\ }\textbf {\bibinfo {volume} {7}},\ \bibinfo {pages}
  {041032} (\bibinfo {year} {2017})}\BibitemShut {NoStop}%
\bibitem [{\citenamefont {Cho}\ \emph {et~al.}(2014)\citenamefont {Cho},
  \citenamefont {You},\ and\ \citenamefont {Fradkin}}]{Cho:WZJain1}%
  \BibitemOpen
  \bibfield  {author} {\bibinfo {author} {\bibfnamefont {G.~Y.}\ \bibnamefont
  {Cho}}, \bibinfo {author} {\bibfnamefont {Y.}~\bibnamefont {You}},\ and\
  \bibinfo {author} {\bibfnamefont {E.}~\bibnamefont {Fradkin}},\ }\bibfield
  {title} {\bibinfo {title} {{Geometry of fractional quantum Hall fluids}},\
  }\href {https://doi.org/10.1103/PhysRevB.90.115139} {\bibfield  {journal}
  {\bibinfo  {journal} {Phys. Rev. B}\ }\textbf {\bibinfo {volume} {90}},\
  \bibinfo {pages} {115139} (\bibinfo {year} {2014})}\BibitemShut {NoStop}%
\bibitem [{\citenamefont {Gromov}\ \emph {et~al.}(2015)\citenamefont {Gromov},
  \citenamefont {Cho}, \citenamefont {You}, \citenamefont {Abanov},\ and\
  \citenamefont {Fradkin}}]{Gromov:WZJain2}%
  \BibitemOpen
  \bibfield  {author} {\bibinfo {author} {\bibfnamefont {A.}~\bibnamefont
  {Gromov}}, \bibinfo {author} {\bibfnamefont {G.~Y.}\ \bibnamefont {Cho}},
  \bibinfo {author} {\bibfnamefont {Y.}~\bibnamefont {You}}, \bibinfo {author}
  {\bibfnamefont {A.~G.}\ \bibnamefont {Abanov}},\ and\ \bibinfo {author}
  {\bibfnamefont {E.}~\bibnamefont {Fradkin}},\ }\bibfield  {title} {\bibinfo
  {title} {{Framing Anomaly in the Effective Theory of the Fractional Quantum
  Hall Effect}},\ }\href {https://doi.org/10.1103/PhysRevLett.114.016805}
  {\bibfield  {journal} {\bibinfo  {journal} {Phys. Rev. Lett.}\ }\textbf
  {\bibinfo {volume} {114}},\ \bibinfo {pages} {016805} (\bibinfo {year}
  {2015})}\BibitemShut {NoStop}%
\bibitem [{\citenamefont {Hoyos}\ and\ \citenamefont {Son}(2012)}]{Son-Hoyos}%
  \BibitemOpen
  \bibfield  {author} {\bibinfo {author} {\bibfnamefont {C.}~\bibnamefont
  {Hoyos}}\ and\ \bibinfo {author} {\bibfnamefont {D.~T.}\ \bibnamefont
  {Son}},\ }\bibfield  {title} {\bibinfo {title} {{Hall Viscosity and
  Electromagnetic Response}},\ }\href
  {https://doi.org/10.1103/PhysRevLett.108.066805} {\bibfield  {journal}
  {\bibinfo  {journal} {Phys. Rev. Lett.}\ }\textbf {\bibinfo {volume} {108}},\
  \bibinfo {pages} {066805} (\bibinfo {year} {2012})}\BibitemShut {NoStop}%
\bibitem [{\citenamefont {Nguyen}\ and\ \citenamefont
  {Son}(2018)}]{Nguyen2018b}%
  \BibitemOpen
  \bibfield  {author} {\bibinfo {author} {\bibfnamefont {D.~X.}\ \bibnamefont
  {Nguyen}}\ and\ \bibinfo {author} {\bibfnamefont {D.~T.}\ \bibnamefont
  {Son}},\ }\bibfield  {title} {\bibinfo {title} {{Algebraic approach to
  fractional quantum Hall effect}},\ }\href
  {https://doi.org/10.1103/physrevb.98.241110} {\bibfield  {journal} {\bibinfo
  {journal} {Phys. Rev. B}\ }\textbf {\bibinfo {volume} {98}},\ \bibinfo
  {pages} {241110} (\bibinfo {year} {2018})}\BibitemShut {NoStop}%
\bibitem [{\citenamefont {Haldane}(2011)}]{Haldane2011}%
  \BibitemOpen
  \bibfield  {author} {\bibinfo {author} {\bibfnamefont {F.~D.~M.}\
  \bibnamefont {Haldane}},\ }\bibfield  {title} {\bibinfo {title} {{Geometrical
  Description of the Fractional Quantum Hall Effect}},\ }\href
  {https://doi.org/10.1103/physrevlett.107.116801} {\bibfield  {journal}
  {\bibinfo  {journal} {Phys. Rev. Lett.}\ }\textbf {\bibinfo {volume} {107}},\
  \bibinfo {pages} {116801} (\bibinfo {year} {2011})}\BibitemShut {NoStop}%
\bibitem [{Note2()}]{Note2}%
  \BibitemOpen
  \bibinfo {note} {The Land\'e factor is introduced for convenience and does
  not change physical quantities computed in this paper, for example, the
  static structure factor.}\BibitemShut {Stop}%
\bibitem [{\citenamefont {Son}()}]{Son2013}%
  \BibitemOpen
  \bibfield  {author} {\bibinfo {author} {\bibfnamefont {D.~T.}\ \bibnamefont
  {Son}},\ }\bibfield  {title} {\bibinfo {title} {{Newton-Cartan Geometry and
  the Quantum Hall Effect}},\ }\href@noop {} {\ }\Eprint
  {https://arxiv.org/abs/1306.0638} {arXiv:1306.0638} \BibitemShut {NoStop}%
\bibitem [{\citenamefont {Geracie}\ \emph {et~al.}(2015)\citenamefont
  {Geracie}, \citenamefont {Son}, \citenamefont {Wu},\ and\ \citenamefont
  {Wu}}]{Geracie:2015}%
  \BibitemOpen
  \bibfield  {author} {\bibinfo {author} {\bibfnamefont {M.}~\bibnamefont
  {Geracie}}, \bibinfo {author} {\bibfnamefont {D.~T.}\ \bibnamefont {Son}},
  \bibinfo {author} {\bibfnamefont {C.}~\bibnamefont {Wu}},\ and\ \bibinfo
  {author} {\bibfnamefont {S.-F.}\ \bibnamefont {Wu}},\ }\bibfield  {title}
  {\bibinfo {title} {{Spacetime symmetries of the quantum Hall effect}},\
  }\href {https://doi.org/10.1103/physrevd.91.045030} {\bibfield  {journal}
  {\bibinfo  {journal} {Phys. Rev. D}\ }\textbf {\bibinfo {volume} {91}},\
  \bibinfo {pages} {045030} (\bibinfo {year} {2015})}\BibitemShut {NoStop}%
\bibitem [{\citenamefont {Prabhu}\ and\ \citenamefont
  {Roberts}(2017)}]{prabhu2017electrons}%
  \BibitemOpen
  \bibfield  {author} {\bibinfo {author} {\bibfnamefont {K.}~\bibnamefont
  {Prabhu}}\ and\ \bibinfo {author} {\bibfnamefont {M.~M.}\ \bibnamefont
  {Roberts}},\ }\href@noop {} {\bibinfo {title} {{Electrons and composite Dirac
  fermions in the lowest Landau level}}} (\bibinfo {year} {2017}),\ \Eprint
  {https://arxiv.org/abs/1709.02814} {arXiv:1709.02814} \BibitemShut {NoStop}%
\bibitem [{Note3()}]{Note3}%
  \BibitemOpen
  \bibinfo {note} {We take $\protect \mathfrak {g}=2$ only for convenience in
  which the mathematical treatment of the lowest Landau level limit is
  simplified. A different value of $\protect \mathfrak {g}$ only modifies
  chemical potential by a constant value \cite {Geracie:2015} that doesn't
  change the physical conclusions in this paper.}\BibitemShut {Stop}%
\bibitem [{\citenamefont {Nguyen}\ \emph
  {et~al.}(2018{\natexlab{b}})\citenamefont {Nguyen}, \citenamefont {Gromov},\
  and\ \citenamefont {Son}}]{Nguyen2018}%
  \BibitemOpen
  \bibfield  {author} {\bibinfo {author} {\bibfnamefont {D.~X.}\ \bibnamefont
  {Nguyen}}, \bibinfo {author} {\bibfnamefont {A.}~\bibnamefont {Gromov}},\
  and\ \bibinfo {author} {\bibfnamefont {D.~T.}\ \bibnamefont {Son}},\
  }\bibfield  {title} {\bibinfo {title} {{Fractional quantum Hall systems near
  nematicity: Bimetric theory, composite fermions, and Dirac brackets}},\
  }\href {https://doi.org/10.1103/physrevb.97.195103} {\bibfield  {journal}
  {\bibinfo  {journal} {Phys. Rev. B}\ }\textbf {\bibinfo {volume} {97}},\
  \bibinfo {pages} {195103} (\bibinfo {year} {2018}{\natexlab{b}})}\BibitemShut
  {NoStop}%
\bibitem [{Note4()}]{Note4}%
  \BibitemOpen
  \bibinfo {note} {Which was also found in the IQH experiment of graphene after
  dividing the experimental result of $\sigma _{xy}$ by the total number of
  spins and valleys \cite {Gusynin:2005}.}\BibitemShut {Stop}%
\bibitem [{\citenamefont {Golkar}\ \emph {et~al.}(2014)\citenamefont {Golkar},
  \citenamefont {Roberts},\ and\ \citenamefont {Son}}]{Golkar:2014}%
  \BibitemOpen
  \bibfield  {author} {\bibinfo {author} {\bibfnamefont {S.}~\bibnamefont
  {Golkar}}, \bibinfo {author} {\bibfnamefont {M.~M.}\ \bibnamefont
  {Roberts}},\ and\ \bibinfo {author} {\bibfnamefont {D.~T.}\ \bibnamefont
  {Son}},\ }\bibfield  {title} {\bibinfo {title} {{Effective field theory of
  relativistic quantum Hall systems}},\ }\href
  {https://doi.org/10.1007/jhep12(2014)138} {\bibfield  {journal} {\bibinfo
  {journal} {J. High Energy Phys.}\ }\textbf {\bibinfo {volume} {2014}}\bibinfo
   {number} { (12)},\ \bibinfo {pages} {138}}\BibitemShut {NoStop}%
\bibitem [{Note5()}]{Note5}%
  \BibitemOpen
\bibfield  {number} {  }\bibinfo {note} {The similar effective action of
  non-relativistic IQH states can be found in \cite {GromovIQH}}\BibitemShut
  {NoStop}%
\bibitem [{Note6()}]{Note6}%
  \BibitemOpen
  \bibinfo {note} {Note that, due to the additional dipole term, one need to
  shift $a_0\rightarrow a_0+\protect \frac {1}{2}\protect \frac {k_F}{v_F}v_i
  v^i$ and $a_i \rightarrow a_i -\protect \frac {k_F}{v_F}v_i$ as explained in
  \cite {Nguyen2018a}. However, the extra terms are higher derivative and do
  not modify the results of this paper.}\BibitemShut {Stop}%
\bibitem [{Note7()}]{Note7}%
  \BibitemOpen
  \bibinfo {note} {Due to the definition of $\omega _0$, there is an extra
  contribution to the charge density with the form $\protect \frac {1}{\protect
  \sqrt {g}}\partial _i \partial ^i (\protect \cdots )$. However, this term
  will not contribute to the Wen-Zee shift on a closed manifold.}\BibitemShut
  {Stop}%
\bibitem [{Note8()}]{Note8}%
  \BibitemOpen
  \bibinfo {note} {Note the definition of $\protect \tilde {A}$ in Eq.~\protect
  \textup {\hbox {\mathsurround \z@ \protect \normalfont (\ignorespaces \ref
  {eq:Amu}\unskip \@@italiccorr )}}}\BibitemShut {NoStop}%
\bibitem [{\citenamefont {Wen}\ and\ \citenamefont {Zee}(1992)}]{Wen-Zee}%
  \BibitemOpen
  \bibfield  {author} {\bibinfo {author} {\bibfnamefont {X.~G.}\ \bibnamefont
  {Wen}}\ and\ \bibinfo {author} {\bibfnamefont {A.}~\bibnamefont {Zee}},\
  }\bibfield  {title} {\bibinfo {title} {{Shift and spin vector: New
  topological quantum numbers for the Hall fluids}},\ }\href
  {https://doi.org/10.1103/PhysRevLett.69.953} {\bibfield  {journal} {\bibinfo
  {journal} {Phys. Rev. Lett.}\ }\textbf {\bibinfo {volume} {69}},\ \bibinfo
  {pages} {953} (\bibinfo {year} {1992})}\BibitemShut {NoStop}%
\bibitem [{\citenamefont {Girvin}\ \emph {et~al.}(1986)\citenamefont {Girvin},
  \citenamefont {MacDonald},\ and\ \citenamefont {Platzman}}]{GMP:1986}%
  \BibitemOpen
  \bibfield  {author} {\bibinfo {author} {\bibfnamefont {S.~M.}\ \bibnamefont
  {Girvin}}, \bibinfo {author} {\bibfnamefont {A.~H.}\ \bibnamefont
  {MacDonald}},\ and\ \bibinfo {author} {\bibfnamefont {P.~M.}\ \bibnamefont
  {Platzman}},\ }\bibfield  {title} {\bibinfo {title} {{Magneto-roton theory of
  collective excitations in the fractional quantum Hall effect}},\ }\href
  {https://doi.org/10.1103/PhysRevB.33.2481} {\bibfield  {journal} {\bibinfo
  {journal} {Phys. Rev. B}\ }\textbf {\bibinfo {volume} {33}},\ \bibinfo
  {pages} {2481} (\bibinfo {year} {1986})}\BibitemShut {NoStop}%
\bibitem [{\citenamefont {Avan}\ and\ \citenamefont
  {Jevicki}(1991)}]{AVAN:1991}%
  \BibitemOpen
  \bibfield  {author} {\bibinfo {author} {\bibfnamefont {J.}~\bibnamefont
  {Avan}}\ and\ \bibinfo {author} {\bibfnamefont {A.}~\bibnamefont {Jevicki}},\
  }\bibfield  {title} {\bibinfo {title} {{Classical integrability and higher
  symmetries of collective string field theory}},\ }\href
  {https://doi.org/https://doi.org/10.1016/0370-2693(91)90740-H} {\bibfield
  {journal} {\bibinfo  {journal} {Phys. Lett. B}\ }\textbf {\bibinfo {volume}
  {266}},\ \bibinfo {pages} {35} (\bibinfo {year} {1991})}\BibitemShut
  {NoStop}%
\bibitem [{\citenamefont {Gerasimov}\ \emph {et~al.}(1991)\citenamefont
  {Gerasimov}, \citenamefont {Marshakov}, \citenamefont {Mironov},
  \citenamefont {Morozov},\ and\ \citenamefont {Orlov}}]{GERASIMOV:1991}%
  \BibitemOpen
  \bibfield  {author} {\bibinfo {author} {\bibfnamefont {A.}~\bibnamefont
  {Gerasimov}}, \bibinfo {author} {\bibfnamefont {A.}~\bibnamefont
  {Marshakov}}, \bibinfo {author} {\bibfnamefont {A.}~\bibnamefont {Mironov}},
  \bibinfo {author} {\bibfnamefont {A.}~\bibnamefont {Morozov}},\ and\ \bibinfo
  {author} {\bibfnamefont {A.}~\bibnamefont {Orlov}},\ }\bibfield  {title}
  {\bibinfo {title} {{Matrix models of two-dimensional gravity and Toda
  theory}},\ }\href
  {https://doi.org/https://doi.org/10.1016/0550-3213(91)90482-D} {\bibfield
  {journal} {\bibinfo  {journal} {Nucl. Phys. B}\ }\textbf {\bibinfo {volume}
  {357}},\ \bibinfo {pages} {565} (\bibinfo {year} {1991})}\BibitemShut
  {NoStop}%
\bibitem [{\citenamefont {Iso}\ \emph {et~al.}(1992)\citenamefont {Iso},
  \citenamefont {Karabali},\ and\ \citenamefont {Sakita}}]{ISO:1992}%
  \BibitemOpen
  \bibfield  {author} {\bibinfo {author} {\bibfnamefont {S.}~\bibnamefont
  {Iso}}, \bibinfo {author} {\bibfnamefont {D.}~\bibnamefont {Karabali}},\ and\
  \bibinfo {author} {\bibfnamefont {B.}~\bibnamefont {Sakita}},\ }\bibfield
  {title} {\bibinfo {title} {{Fermions in the lowest Landau level.
  Bosonization, $W_{\infty}$ algebra, droplets, chiral bosons}},\ }\href
  {https://doi.org/https://doi.org/10.1016/0370-2693(92)90816-M} {\bibfield
  {journal} {\bibinfo  {journal} {Phys. Lett. B}\ }\textbf {\bibinfo {volume}
  {296}},\ \bibinfo {pages} {143} (\bibinfo {year} {1992})}\BibitemShut
  {NoStop}%
\bibitem [{\citenamefont {Golkar}\ \emph {et~al.}(2016)\citenamefont {Golkar},
  \citenamefont {Nguyen}, \citenamefont {Roberts},\ and\ \citenamefont
  {Son}}]{Golkar2016}%
  \BibitemOpen
  \bibfield  {author} {\bibinfo {author} {\bibfnamefont {S.}~\bibnamefont
  {Golkar}}, \bibinfo {author} {\bibfnamefont {D.~X.}\ \bibnamefont {Nguyen}},
  \bibinfo {author} {\bibfnamefont {M.~M.}\ \bibnamefont {Roberts}},\ and\
  \bibinfo {author} {\bibfnamefont {D.~T.}\ \bibnamefont {Son}},\ }\bibfield
  {title} {\bibinfo {title} {{Higher-Spin Theory of the Magnetorotons}},\
  }\href {https://doi.org/10.1103/physrevlett.117.216403} {\bibfield  {journal}
  {\bibinfo  {journal} {Phys. Rev. Lett.}\ }\textbf {\bibinfo {volume} {117}},\
  \bibinfo {pages} {216403} (\bibinfo {year} {2016})}\BibitemShut {NoStop}%
\bibitem [{Note9()}]{Note9}%
  \BibitemOpen
  \bibinfo {note} {It has been shown that the the semiclassical approach gives
  the same results of electromagnetic response as the random phase
  approximation calculation~\cite {NguyenLLEM}.}\BibitemShut {Stop}%
\bibitem [{\citenamefont {Luttinger}(1960)}]{Luttinger:1960}%
  \BibitemOpen
  \bibfield  {author} {\bibinfo {author} {\bibfnamefont {J.~M.}\ \bibnamefont
  {Luttinger}},\ }\bibfield  {title} {\bibinfo {title} {Fermi surface and some
  simple equilibrium properties of a system of interacting fermions},\ }\href
  {https://doi.org/10.1103/PhysRev.119.1153} {\bibfield  {journal} {\bibinfo
  {journal} {Phys. Rev.}\ }\textbf {\bibinfo {volume} {119}},\ \bibinfo {pages}
  {1153} (\bibinfo {year} {1960})}\BibitemShut {NoStop}%
\bibitem [{\citenamefont {Haldane}(1994)}]{Haldane:1994}%
  \BibitemOpen
  \bibfield  {author} {\bibinfo {author} {\bibfnamefont {F.~D.~M.}\
  \bibnamefont {Haldane}},\ }\bibfield  {title} {\bibinfo {title} {Luttinger's
  theorem and bosonization of the fermi surface},\ }in\ \href@noop {} {\emph
  {\bibinfo {booktitle} {Perspectives in Many-Particle Physics}}},\ \bibinfo
  {editor} {edited by\ \bibinfo {editor} {\bibfnamefont {R.~A.}\ \bibnamefont
  {Broglia}}, \bibinfo {editor} {\bibfnamefont {J.~R.}\ \bibnamefont
  {Schrieffer}},\ and\ \bibinfo {editor} {\bibfnamefont {P.~F.}\ \bibnamefont
  {Bortignon}}}\ (\bibinfo  {publisher} {North-Holland},\ \bibinfo {address}
  {Amsterdam},\ \bibinfo {year} {1994})\ pp.\ \bibinfo {pages} {5--30},\
  \Eprint {https://arxiv.org/abs/cond-mat/0505529} {cond-mat/0505529}
  \BibitemShut {NoStop}%
\bibitem [{\citenamefont {Nguyen}\ and\ \citenamefont
  {Gromov}(2017)}]{NguyenLLEM}%
  \BibitemOpen
  \bibfield  {author} {\bibinfo {author} {\bibfnamefont {D.~X.}\ \bibnamefont
  {Nguyen}}\ and\ \bibinfo {author} {\bibfnamefont {A.}~\bibnamefont
  {Gromov}},\ }\bibfield  {title} {\bibinfo {title} {{Exact electromagnetic
  response of Landau level electrons}},\ }\href
  {https://doi.org/10.1103/PhysRevB.95.085151} {\bibfield  {journal} {\bibinfo
  {journal} {Phys. Rev. B}\ }\textbf {\bibinfo {volume} {95}},\ \bibinfo
  {pages} {085151} (\bibinfo {year} {2017})}\BibitemShut {NoStop}%
\bibitem [{Note10()}]{Note10}%
  \BibitemOpen
  \bibinfo {note} {One can derive the same equation of motion using the
  Boltzmann kinetic equation \cite {NguyenLLEM,Nguyen2018a}.}\BibitemShut
  {Stop}%
\bibitem [{Note11()}]{Note11}%
  \BibitemOpen
  \bibinfo {note} {The mixed correlation $\langle \rho ^{(1)}(-\protect \mathbf
  {q})\rho ^{(2)}(\protect \mathbf {q}) \rangle $ vanishes.}\BibitemShut
  {Stop}%
\bibitem [{\citenamefont {Nguyen}\ \emph {et~al.}(2014)\citenamefont {Nguyen},
  \citenamefont {Son},\ and\ \citenamefont {Wu}}]{Nguyen:2014LLL}%
  \BibitemOpen
  \bibfield  {author} {\bibinfo {author} {\bibfnamefont {D.~X.}\ \bibnamefont
  {Nguyen}}, \bibinfo {author} {\bibfnamefont {D.~T.}\ \bibnamefont {Son}},\
  and\ \bibinfo {author} {\bibfnamefont {C.}~\bibnamefont {Wu}},\ }\href@noop
  {} {\bibinfo {title} {{Lowest Landau Level Stress Tensor and Structure Factor
  of Trial Quantum Hall Wave Functions}}} (\bibinfo {year} {2014}),\ \Eprint
  {https://arxiv.org/abs/1411.3316} {arXiv:1411.3316} \BibitemShut {NoStop}%
\bibitem [{\citenamefont {Nguyen}\ \emph {et~al.}(2017)\citenamefont {Nguyen},
  \citenamefont {Can},\ and\ \citenamefont {Gromov}}]{Nguyen:PH}%
  \BibitemOpen
  \bibfield  {author} {\bibinfo {author} {\bibfnamefont {D.~X.}\ \bibnamefont
  {Nguyen}}, \bibinfo {author} {\bibfnamefont {T.}~\bibnamefont {Can}},\ and\
  \bibinfo {author} {\bibfnamefont {A.}~\bibnamefont {Gromov}},\ }\bibfield
  {title} {\bibinfo {title} {Particle-hole duality in the lowest landau
  level},\ }\href {https://doi.org/10.1103/PhysRevLett.118.206602} {\bibfield
  {journal} {\bibinfo  {journal} {Phys. Rev. Lett.}\ }\textbf {\bibinfo
  {volume} {118}},\ \bibinfo {pages} {206602} (\bibinfo {year}
  {2017})}\BibitemShut {NoStop}%
\bibitem [{\citenamefont {Liou}\ \emph {et~al.}(2019)\citenamefont {Liou},
  \citenamefont {Haldane}, \citenamefont {Yang},\ and\ \citenamefont
  {Rezayi}}]{Liou2019}%
  \BibitemOpen
  \bibfield  {author} {\bibinfo {author} {\bibfnamefont {S.-F.}\ \bibnamefont
  {Liou}}, \bibinfo {author} {\bibfnamefont {F.~D.~M.}\ \bibnamefont
  {Haldane}}, \bibinfo {author} {\bibfnamefont {K.}~\bibnamefont {Yang}},\ and\
  \bibinfo {author} {\bibfnamefont {E.~H.}\ \bibnamefont {Rezayi}},\ }\bibfield
   {title} {\bibinfo {title} {{Chiral Gravitons in Fractional Quantum Hall
  Liquids}},\ }\href {https://doi.org/10.1103/PhysRevLett.123.146801}
  {\bibfield  {journal} {\bibinfo  {journal} {Phys. Rev. Lett.}\ }\textbf
  {\bibinfo {volume} {123}},\ \bibinfo {pages} {146801} (\bibinfo {year}
  {2019})}\BibitemShut {NoStop}%
\bibitem [{\citenamefont {Haldane}\ \emph {et~al.}(2021)\citenamefont
  {Haldane}, \citenamefont {Rezayi},\ and\ \citenamefont {Yang}}]{haldane2021}%
  \BibitemOpen
  \bibfield  {author} {\bibinfo {author} {\bibfnamefont {F.~D.~M.}\
  \bibnamefont {Haldane}}, \bibinfo {author} {\bibfnamefont {E.~H.}\
  \bibnamefont {Rezayi}},\ and\ \bibinfo {author} {\bibfnamefont
  {K.}~\bibnamefont {Yang}},\ }\bibfield  {title} {\bibinfo {title} {{Graviton
  Chirality and Topological Order in the Half-filled Landau Level}},\
  }\href@noop {} {\  (\bibinfo {year} {2021})},\ \Eprint
  {https://arxiv.org/abs/2103.11019} {arXiv:2103.11019} \BibitemShut {NoStop}%
\bibitem [{\citenamefont {Nguyen}\ and\ \citenamefont {Son}(2021)}]{Raman}%
  \BibitemOpen
  \bibfield  {author} {\bibinfo {author} {\bibfnamefont {D.~X.}\ \bibnamefont
  {Nguyen}}\ and\ \bibinfo {author} {\bibfnamefont {D.~T.}\ \bibnamefont
  {Son}},\ }\bibfield  {title} {\bibinfo {title} {{Probing the spin structure
  of the fractional quantum Hall magnetoroton with polarized Raman
  scattering}},\ }\href {https://doi.org/10.1103/physrevresearch.3.023040}
  {\bibfield  {journal} {\bibinfo  {journal} {Phys. Rev. Res.}\ }\textbf
  {\bibinfo {volume} {3}},\ \bibinfo {pages} {023040} (\bibinfo {year}
  {2021})}\BibitemShut {NoStop}%
\bibitem [{Note12()}]{Note12}%
  \BibitemOpen
  \bibinfo {note} {The upper limit $\infty $ here means a frequency which is
  much higher than the Coulomb gap but also smaller the the cyclotron
  gap.}\BibitemShut {Stop}%
\bibitem [{Note13()}]{Note13}%
  \BibitemOpen
  \bibinfo {note} {The extra terms (second line in the equation \protect
  \textup {\hbox {\mathsurround \z@ \protect \normalfont (\ignorespaces \ref
  {eq:comPi}\unskip \@@italiccorr )}}) disappear when we compute the curl's in
  $[\rho ^{(1)}(\protect \mathbf {x}), \protect \, \rho ^{(1)}(\protect \mathbf
  {y})]$.}\BibitemShut {Stop}%
\bibitem [{\citenamefont {Gusynin}\ and\ \citenamefont
  {Sharapov}(2005)}]{Gusynin:2005}%
  \BibitemOpen
  \bibfield  {author} {\bibinfo {author} {\bibfnamefont {V.~P.}\ \bibnamefont
  {Gusynin}}\ and\ \bibinfo {author} {\bibfnamefont {S.~G.}\ \bibnamefont
  {Sharapov}},\ }\bibfield  {title} {\bibinfo {title} {{Unconventional Integer
  Quantum Hall Effect in Graphene}},\ }\href
  {https://doi.org/10.1103/physrevlett.95.146801} {\bibfield  {journal}
  {\bibinfo  {journal} {Phys. Rev. Lett.}\ }\textbf {\bibinfo {volume} {95}},\
  \bibinfo {pages} {146801} (\bibinfo {year} {2005})}\BibitemShut {NoStop}%
\bibitem [{\citenamefont {Abanov}\ and\ \citenamefont
  {Gromov}(2014)}]{GromovIQH}%
  \BibitemOpen
  \bibfield  {author} {\bibinfo {author} {\bibfnamefont {A.~G.}\ \bibnamefont
  {Abanov}}\ and\ \bibinfo {author} {\bibfnamefont {A.}~\bibnamefont
  {Gromov}},\ }\bibfield  {title} {\bibinfo {title} {Electromagnetic and
  gravitational responses of two-dimensional noninteracting electrons in a
  background magnetic field},\ }\href
  {https://doi.org/10.1103/PhysRevB.90.014435} {\bibfield  {journal} {\bibinfo
  {journal} {Phys. Rev. B}\ }\textbf {\bibinfo {volume} {90}},\ \bibinfo
  {pages} {014435} (\bibinfo {year} {2014})}\BibitemShut {NoStop}%
\end{thebibliography}%
	
\end{document}